\documentclass{acmart}

\usepackage{verbatim}
\usepackage{adjustbox}
\usepackage{fancyvrb}
\usepackage{versions}
\excludeversion{TODO}\excludeversion{MAYB}
\excludeversion{JeremySays}\excludeversion{CristobalSays}\excludeversion{FabianSays}
\excludeversion{INUTILE}
\includeversion{LONG}\excludeversion{SHORT}
\includeversion{VLONG}
\includeversion{NONANONYMOUS}\excludeversion{ANONYMOUS}
\includeversion{HEADPARAGRAPH}
\includeversion{BRIDGEPARAGRAPH}

\providecommand\contribname{Contributions}%
\specialcomment{contrib}{%
  \begingroup
  \section*{\contribname}
  \phantomsection\addcontentsline{toc}{section}{\contribname}
}{%
  \endgroup
}

\AtBeginDocument{%
  \providecommand\BibTeX{{%
    \normalfont B\kern-0.5em{\scshape i\kern-0.25em b}\kern-0.8em\TeX}}}

\begin{document}

\title%
[Measuring Discrimination Abilities (...) Through a Digital Life Enrichment Application]%
{Measuring Discrimination Abilities of Monk Parakeets Between Discreet and Continuous Quantities Through a Digital Life Enrichment Application}

\author{Jérémy Barbay}
\email{jeremy@barbay.cl}
\orcid{0000-0002-3392-8353}
\author{Fabi\'an Jaña Ubal}
\email{fmju96@gmail.com}
\author{Cristóbal Sepulveda Álvarez}
\email{c.sepulveda.a23@gmail.com}
\affiliation{%
  \institution{\\
  Departamento de Ciencias de la Computación (DCC), 
  Universidad de Chile}
  \streetaddress{Avenida Beauchef 851}
  \city{Santiago}
  \state{Region Metropolitana}
  \country{Chile}
  \postcode{8370448}
}
\renewcommand{\shortauthors}{Barbay et al.}

\begin{abstract}
Ain et al. measured three African Grey (\emph{Psittacus erithacus}) parrot's discrimination abilities between discreet and continuous quantities.  Some features of their experimental protocol make it difficult to apply to other subjects and/or species without introducing a risk for some bias, as subjects could read cues from the experimenter (even though the study's subjects probably did not).  Can digital life enrichment techniques permit us to replicate their results with other species with less risk for experimental bias, with a better precision, and at lower cost?  Inspired by previous informal digital life enrichment experiments with parrots, we designed and tested a web application to digitally replicate and extend Ain et al.'s experimental setup.  We were able to obtain similar results to theirs for two individuals from a distinct species, Monk Parakeets (\emph{Myiopsitta Monachus}), with increased guarantees against potential experimental biases, in a way which should allow to replicate such experiments at larger scale and at a much lower cost.
\end{abstract}



\begin{teaserfigure}
\begin{minipage}[b]{.45\textwidth}
  \includegraphics[width=\textwidth]{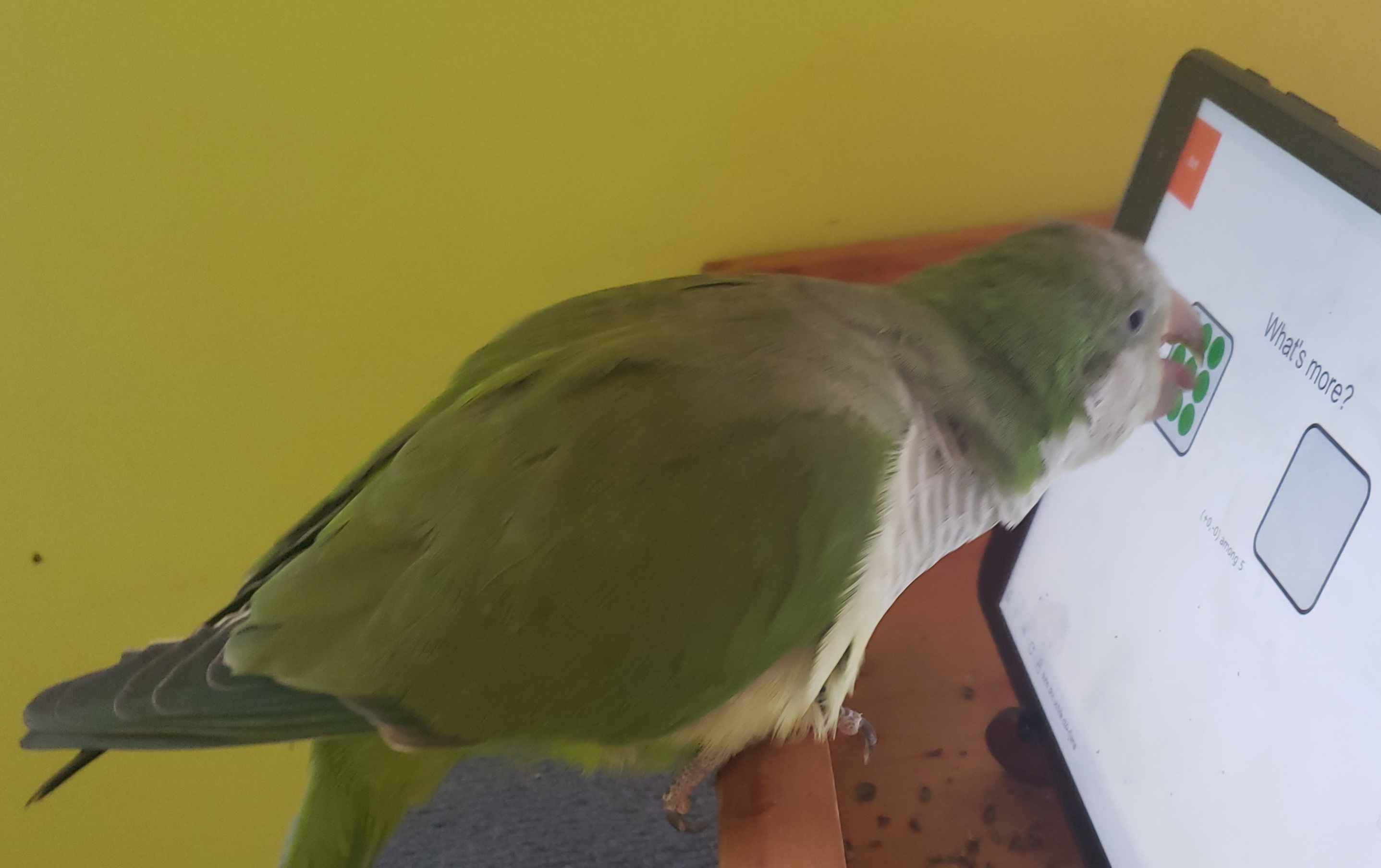}
  \caption{The Monk Parakeet Tina selecting the largest value out of two displayed, in heap mode.}
  \Description{A monk parakeet in front of the touch screen of a grey smart phone "Galaxy Note 9" reaches to select the largest disk out of four, below the title "What's more?".}
  \label{fig:teaser2}
  \end{minipage}
  \hfill
\begin{minipage}[b]{.45\textwidth}
  \includegraphics[width=\textwidth]{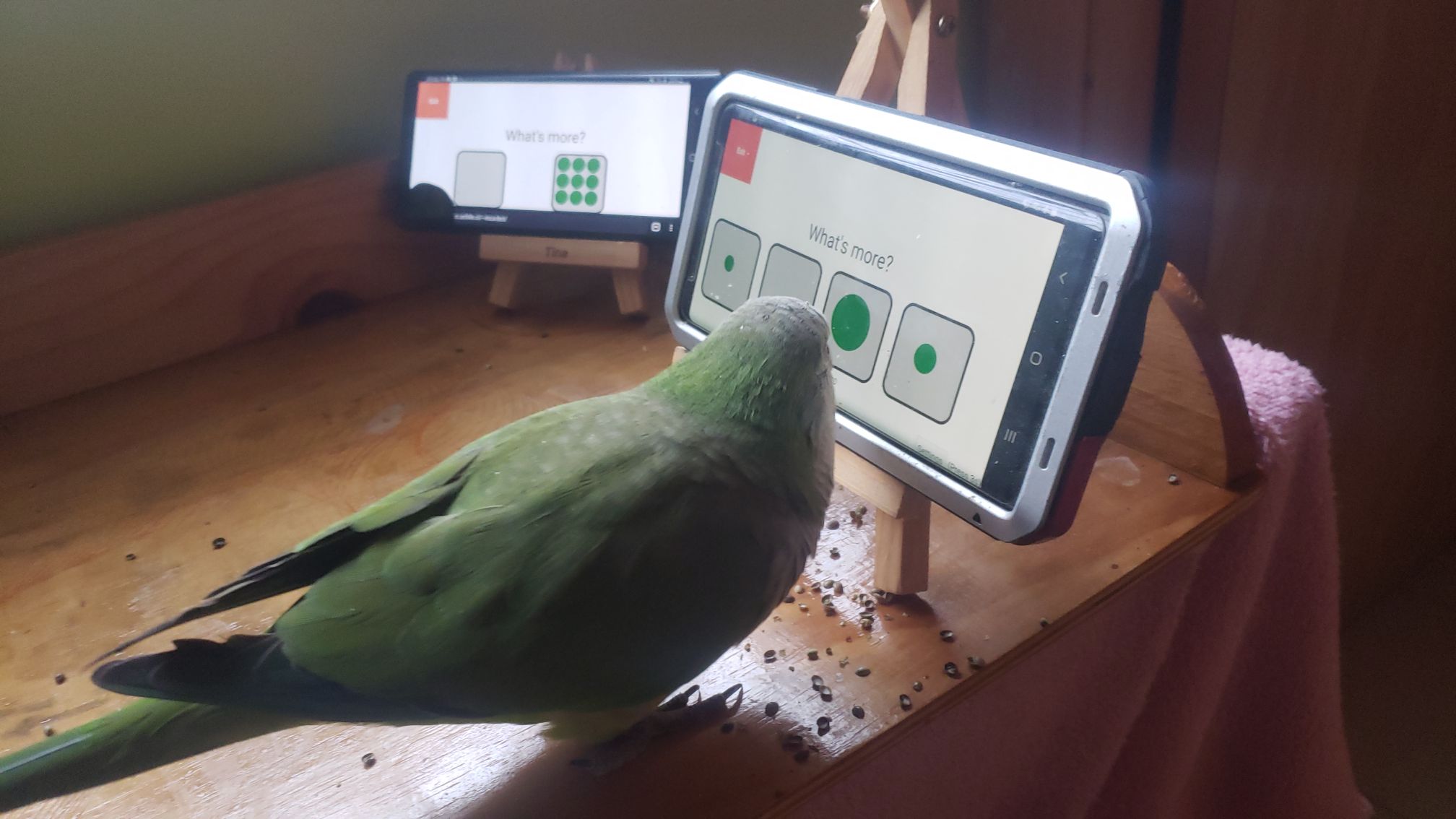}
  \caption{The Monk Parakeet Lorenzo selecting the largest value out of four displayed, in disk mode.}
  \Description{A monk parakeet in front of the touch screen of a grey smart phone "Galaxy Note 9" reaches to select the largest disk out of four, below the title "What's more?".}
  \label{fig:teaser3}
  \end{minipage}
\end{teaserfigure}

\maketitle

\section{Introduction}
\label{sec:orgb1eab2a}
\citet{2008-AC-TheDiscriminationOfDiscreteAndContinuousAmountsInAfricanGreyParrots-AlAinGiretGrandKreutzerBovet} measured the discrimination abilities between discrete and continuous quantities of three  African Grey parrots (\emph{Psittacus erithacus}), showing that their accuracy in choosing between two small quantities was inversely correlated with the ratio between the difference between the two quantities and the largest quantity.

Generalizing the experimental protocol described and implemented by \citet{2008-AC-TheDiscriminationOfDiscreteAndContinuousAmountsInAfricanGreyParrots-AlAinGiretGrandKreutzerBovet} to other subjects or species present some difficulties.
The fact that the experimenter knows which answer was expected from the subjects is not an issue in their study because it was previously verified that the three subjects were unable to read such cues from human experimenters, but it means that the replication of such protocol is limited to individuals (from the same or from other species) which inability to read cues has been previously demonstrated. Beyond such a weakness, the cost of the experimental set-up and of the analysis of the video recordings of the experiments reduces the probability that such a protocol will be replicated with other subjects from the same species, or with subjects from the many other species of parrots existing around the world.

Touchscreens have been successfully used for experiments in life enrichment
\cite{2012-ZB-TechnologyAtTheZooTheInfluenceOfATouchscreenComputerOnOrangutansAndZooVisitors-PerdueClayGaalemaMapleStoinski,1994-RST-ZooAndAnimalWelfare-Kohn,2021-EIT-ImprovingEthicalAttitudesToAnimalsWithDigitalTechnologiesTheCaseOfApesAndZoos-CoghlanWebberCarter}
and in Comparative Psychology~\cite{2018-ZB-AReviewOfZooBasedCognitiveResearchUsingTouchsceenInterfaces-EgelkampRoss}, with individuals from various nonhuman species.
Could digital life enrichment techniques allow to replicate \citet{2008-AC-TheDiscriminationOfDiscreteAndContinuousAmountsInAfricanGreyParrots-AlAinGiretGrandKreutzerBovet}'s results at a lower cost, but with a better precision, and less potential experimental bias? Which additional advantages could a digital variant bring?

\begin{figure}
\centering
\begin{minipage}[b]{.3\textwidth}\centering%
  \includegraphics[width=\textwidth]{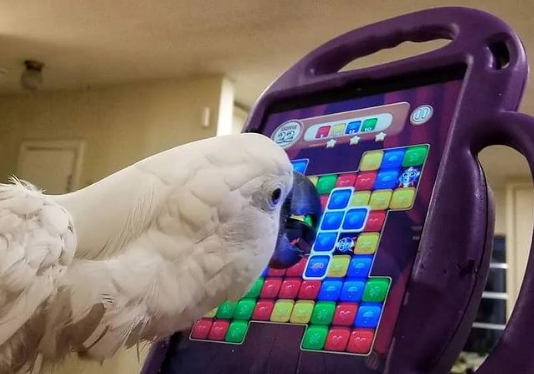}%
  \Description{A Cockatoo parrot playing Candy Crush on a large tablet.}
  \begin{ANONYMOUS}%
  \caption[Candy Crush Example of Digital Life Enrichment]{A Cockatoo
playing the game ``\texttt{Candy Crush}''  (picture used with the authorisation of the author).\label{fig:IsabellePlayingCandyCrush}}%
\end{ANONYMOUS}\begin{NONANONYMOUS}%
\caption[Candy Crush Example of Digital Life Enrichment]{The Cockatoo Isabelle playing the game ``\texttt{Candy Crush}'' under the guidance of Jennifer Cunha from \textsc{Parrot Kindergarten} (picture used with the authorisation of the author).\label{fig:IsabellePlayingCandyCrush}}%
\end{NONANONYMOUS}%
\end{minipage} \hfill\begin{minipage}[b]{.3\textwidth}\centering%
  \includegraphics[width=\textwidth]{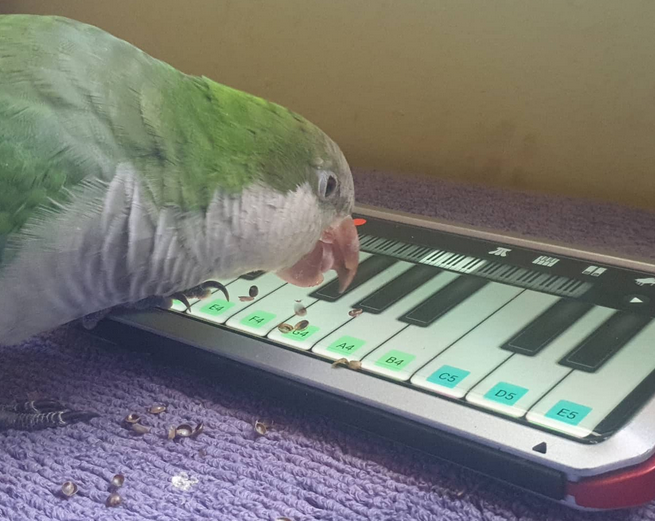}%
  \Description{A Monk Parrot playing a piano app on a grey cell phone lying on the table.}
  \begin{ANONYMOUS}\caption[Piano App Example of Digital Life Enrichment]{Monk Parakeet playing the piano music application ``\texttt{Mini Piano Lite}'' in order to learn to use touchscreen interfaces with a wide active surface.\label{fig:LorenzoPlayingPianoOnTouchScreens}}\end{ANONYMOUS}\begin{NONANONYMOUS}%
  \caption[Piano App Example of Digital Life Enrichment]{The Monk Parakeet Lorenzo playing the piano music application ``\texttt{Mini Piano Lite}'' in order to learn to use touchscreen interfaces with a wide active surface.\label{fig:LorenzoPlayingPianoOnTouchScreens}}\end{NONANONYMOUS}%
\end{minipage}\hfill\begin{minipage}[b]{.3\textwidth}\centering%
  \includegraphics[width=\textwidth]{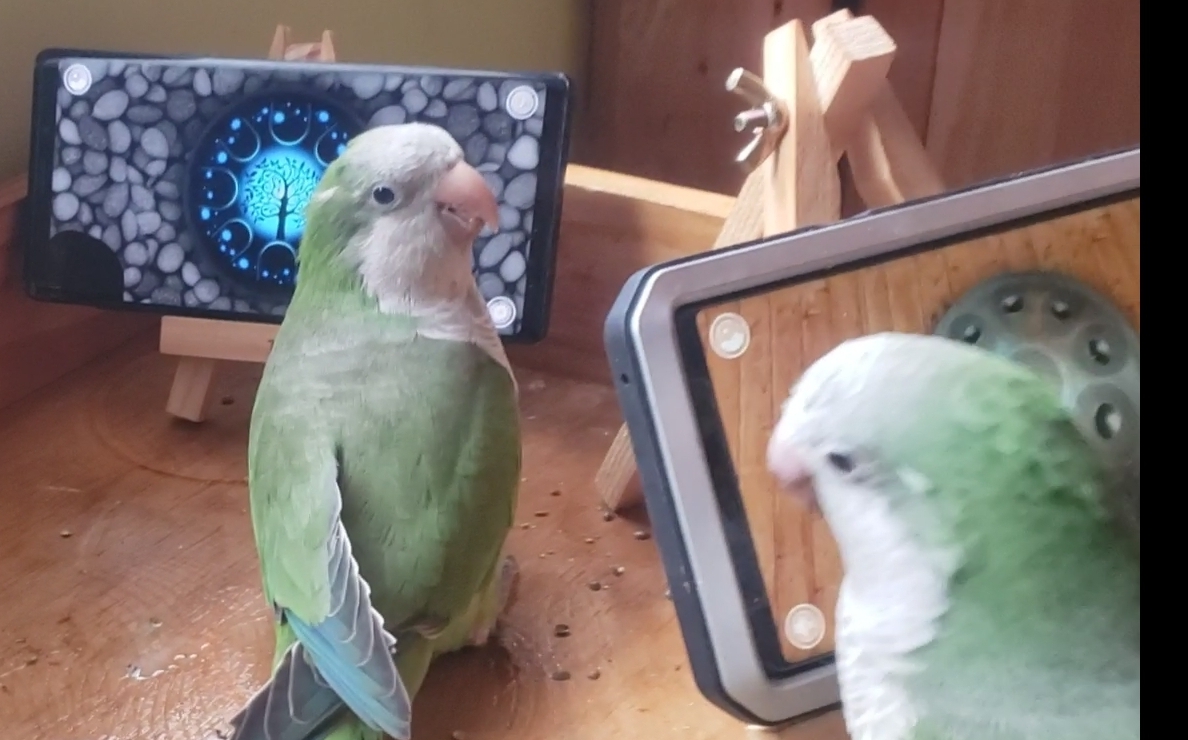}%
  \Description{Two Monk Parrots playing steel drum music on cell phone apps, each with their own device.}
  \begin{ANONYMOUS}\caption[Digital Drum Example of Digital Life Enrichment]{Monk Parakeets playing the steel drum music application ``\texttt{Meditation Drum}''  in order to learn how to properly aim when using touchscreen interfaces.\label{fig:LorenzoAndTinaPlayingSteelDrumOnTouchScreens}}%
  \end{ANONYMOUS}\begin{NONANONYMOUS}%
  \caption[Digital Drum Example of Digital Life Enrichment]{The Monk Parakeets Tina (left) and Lorenzo (right) playing the steel drum music application ``\texttt{Meditation Drum}'' to learn how to aim when using touchscreen interfaces.\label{fig:LorenzoAndTinaPlayingSteelDrumOnTouchScreens}}%
  \end{NONANONYMOUS}%
  \end{minipage}
  \end{figure}

Inspired by previous informal Digital Life Enrichment experiments such as a Cockatoo playing the video game \texttt{Candy Crush} (Figure~\ref{fig:IsabellePlayingCandyCrush}), or Monk Parakeets learning to use touch interfaces by playing music on it (Figures~\ref{fig:LorenzoPlayingPianoOnTouchScreens} and~\ref{fig:LorenzoAndTinaPlayingSteelDrumOnTouchScreens}), we designed, tested and used a web application \begin{NONANONYMOUS} \texttt{InCA-WhatIsMore} \end{NONANONYMOUS} to digitally replicate and extend \citet{2008-AC-TheDiscriminationOfDiscreteAndContinuousAmountsInAfricanGreyParrots-AlAinGiretGrandKreutzerBovet}'s experimental setup.  We obtained similar results to that of Ain et al. for two individuals of a distinct species of parrots, Monk Parakeets (\emph{Myiopsitta Monachus}), using an experimental protocol with increased guarantees against potential experimental biases, at a lower set-up cost, with additional advantages brought by the digital context, such as automatic logging and increased subject's agency.
After describing a selection of concepts and results in the research area of comparative psychology (Section~\ref{sec:cognitiveAbilities}), we describe the application \begin{NONANONYMOUS}\texttt{InCA-WhatIsMore}\end{NONANONYMOUS} (Section~\ref{sec:incaWhatIsMore}), an experimental protocol (including separate development, training and testing phases) based upon it (Section~\ref{sec:experimentations}), an implementation of this protocol and an analysis of its results (Section~\ref{sec:results}), and we conclude with a recapitulation of our results, a discussion of their potential weaknesses and a perspective on future research (Section~\ref{sec:conclusion}).

\section{Nonhuman Cognition}
\label{sec:cognitiveAbilities}

The cognitive abilities of nonhuman animals, traditionally less studied than that of human ones, has been receiving more attention in the last half century. Such studies started with the animals perceived to be ``closest'' to humankind, such as apes, and has spread more recently to birds~\cite{1999-BOOK-TheAlexStudies-Pepperberg,2008-AC-TheDiscriminationOfDiscreteAndContinuousAmountsInAfricanGreyParrots-AlAinGiretGrandKreutzerBovet,2018-ACI-AdvancingCommunicationWithBirdsCanTheyLearnToRead-CunhaClubb}.
\begin{HEADPARAGRAPH}
We describe
\begin{MAYB}
a general overview of some projects and results about the cognitive abilities of some ape and bird species (Section~\ref{sec:ComparativePsychology});
\end{MAYB}
\citet{2008-AC-TheDiscriminationOfDiscreteAndContinuousAmountsInAfricanGreyParrots-AlAinGiretGrandKreutzerBovet}'s study of the discrimination abilities of some parrots (Section~\ref{sec:comparingAbilities});
how devices (analogical and digital) were introduced to nonhumans in order to improve their well being, and often study their abilities at the same time (Section~\ref{sec:lifeEnrichmentApplications}); 
how the distrust in results obtained by improper experimental protocols has plagued scientific research in this area in the past (Section~\ref{sec:experimentalBiases}); and
how some general guiding principles in the  design of experimental protocols permit scientists to avoid such experimental biases (Section~\ref{maskedExperimentalProtocols}).
\end{HEADPARAGRAPH}

\begin{MAYB}
\subsection{Comparative Psychology}
\label{sec:ComparativePsychology}

Comparative psychology refers to the scientific study of the behavior and mental processes of non-human animals (referred to as ``nonhumans'' thereafter), especially as these relate to the phylogenetic history, adaptive significance, and development of behavior in many different species, from insects to primates. 
\citet{2020-FiP-TheComparativePsychologyOfIntelligenceSomeThirtyYearsLater-Pepperberg} describes the history of the field of Comparative Psychology of Intelligence in the last 30 years.

\citet{2020-ExoticsCon-ReadingComprehensionSkillsInAGoffinSCockatoo-Cunha}

\citet{2018-ACI-AdvancingCommunicationWithBirdsCanTheyLearnToRead-CunhaClubb}
\end{MAYB}

\subsection{Discrimination Abilities in African Grey parrots}
\label{sec:comparingAbilities}

\citet{2008-AC-TheDiscriminationOfDiscreteAndContinuousAmountsInAfricanGreyParrots-AlAinGiretGrandKreutzerBovet} tested the discrimination abilities of African Grey (\emph{Psittacus erithacus}) parrots on discrete and continuous amounts. More precisely, they investigated the ability of three African grey parrots to select the largest amount of food between two sets, in two types of experiments. In the first experiment type, the subjects were tested on discrete quantities via the presentation of two quantities of sunflower seeds\begin{LONG} (Deli nature Beyers Belgium)\end{LONG}, between 1,2,3,4 and 5 seeds. In the second experiment type, the subjects were tested on continuous quantities via the presentation of two quantities of parrot formula, with amounts between 0.2,0.4,0.6,0.8 and 1 ml.  For each experiment, the two amounts were presented simultaneously and were visible at the time of choice. Albeit the subjects sometimes failed to choose the largest value, they always performed above chance, their performance improving when the difference between amounts was the greatest.

The experimental setup was completely analogical. A permanent table was set-up in the aviary, and two black pieces of cardboard were used to present food item (sunflower seeds or parrot formula). For each experiment, different amounts of either seeds or parrot formula were placed on each piece of cardboard. The experimenter put the subject for 5 seconds in a position from which they could observe the two sets, then placed them on the table at equal distances from the two sets, letting them chose one set to it while removing the ignored set.  The position of the sets (small and large) was pseudo-randomized: the larger set was never presented more than two times on the same side and was presented as often on the right side as on the left side.

In the experimental setup described by \citet{2008-AC-TheDiscriminationOfDiscreteAndContinuousAmountsInAfricanGreyParrots-AlAinGiretGrandKreutzerBovet}, subjects could eventually read involuntarily cues from the experimenter: even though the experimenter was standing behind the subject, at equal distances from each set, not pointing to it, looking at the subject, aiming to avoid communicating any cue to the subject, the experimenter \emph{knew} where the largest quantity was. While it was not an issue in \citet{2008-AC-TheDiscriminationOfDiscreteAndContinuousAmountsInAfricanGreyParrots-AlAinGiretGrandKreutzerBovet}'s study because the authors demonstrated in a previous study that the subjects were not able to use any gazing cue, the protocol should not be applied as such to other subjects without verifying their inability to read such cues, adding to the cost of implementing such protocol.

Avoiding giving cues to the subject is hard even for a professionally trained experimenter~\cite{2015-BiologicalTheory-Clever_Hans_Alex_the_Parrot_and_Kanzi_WhatCanExceptionalAnimalLearningTeachUsAboutHumanCognitiveEvolution-Trestman}. Requiring either such training or a separate study to insure that the subject cannot read cues from the experimenter restricts the applicability of a protocol to laboratories. 
For example, in the context of \emph{citizen science} projects~\cite{2013-NATURE-CitizenScienceAmateurExperts-Gura}  where non professional experimenters (such as zoo personal or simple citizen) guide the experiments, a masked protocol \begin{LONG}(defined in Section~\ref{maskedExperimentalProtocols})\end{LONG} where the experimenters \emph{ignore} what the correct answer is (because they did not receive the information that the subject did) would be more robust against subjects reading cues from the experimenter.  We describe in section~\ref{sec:incaWhatIsMore} an application allowing for such an alternate experimental setup which, if not exactly equivalent to that of \citet{2008-AC-TheDiscriminationOfDiscreteAndContinuousAmountsInAfricanGreyParrots-AlAinGiretGrandKreutzerBovet} (e.g. the reward is not proportional to the quantity selected), presents the advantage of being ``experimenter-masked'', inspired by some of the life enrichment experiences described in the next section.

\subsection{Life Enrichment and Cognition studies}
\label{sec:lifeEnrichmentApplications}

One can study cognitive abilities of nonhumans through life enrichment activities in general, and through digital ones in particular.
\begin{LONG}
General preoccupation for the welfare of captive nonhumans is at least 150 years old.  \citet{1994-RST-ZooAndAnimalWelfare-Kohn} dates the first legislation about zoo animal welfare to 1876, with the ``Cruelty to Animals Act'' in the ``Criminal Code of Canada''. Since then, the list of duties of such institutions has grown to include not only the basic welfare tenets of adequate feed, water, shelter, sanitation and veterinary care of their nonhuman residents, but also higher level concerns such as the handling and training of the nonhuman residents, their psychological well-being, the design of their enclosures, the preservation of their species, issues of environmental and conservation, and programs to breed captive nonhumans.
\citet{1994-RST-ZooAndAnimalWelfare-Kohn} mentions (in 1994) the ``\emph{emerging field of psychological well-being in captive animals}'', incorporating physical health, normal and captive behavior, and interactions with the enclosure environments and mentioning how environmental enrichment is an important component of this issue. He goes on to list innovations in life enrichment such as specialized toys and puzzle feed boxes (but no digital applications).

Yet, the use of digital applications to measure nonhuman abilities seems to predate 
\citet{1994-RST-ZooAndAnimalWelfare-Kohn}'s report by at least 10 years.
In his discussion of the impact of game-like computerized tasks designed to promote and assess the psychological well-being of captive nonhuman, \citet{2015-ABC-TheFourCsOfPsychologicalWellbeingLessonsFrom3DecadesOfComputerBasedEnvironmentalEnrichment-Washburn} refers to a three decade old history in 2015, placing the beginning of such use sometimes around 1985.
\end{LONG}
In 1990, \citet{1990-BRMIC-TheNASALRCComputerizedTestSystem-RichardsonWashburnHopkinsSavageRumbaughRumbaugh} describe a quite complete Computerized Test System.
\begin{LONG}
They tested their system with a population of rhesus monkeys, but defend its potential as a ``\emph{rich and robust testing environment for the cognitive and neuro-psychological capacities of great apes, rhesus monkeys, mentally retarded and normally developing children, and adults}'', so that subjects from various populations can be tested under comparable conditions in such a way that ``\emph{control is increased over testing conditions}''.
\end{LONG}
They mention that ``\emph{the animals readily started to work even when the reward was a small pellet of chow very similar in composition to the chow just removed from the cage}'', and that ``\emph{the tasks have some motivating or rewarding of their own}''.

Nonhuman subjects seem to enjoy participating in cognitive studies involving game-like digital applications.
\citet{2015-ABC-TheFourCsOfPsychologicalWellbeingLessonsFrom3DecadesOfComputerBasedEnvironmentalEnrichment-Washburn}
describes, among various other anecdotes, how game-like application for apes were developed as early as 1984, and how the subjects ``\emph{chose to work on joystick-based tasks, even though they did not need to perform the game-like tests in order to receive food}'', and ``\emph{opted for computer task activity over other potential activities that were available to them}''.
\begin{LONG}
He goes on to mention how such game-like activities have been used to study various cognitive phenomena such as the ability to learn, memory, attention, perception, categorization, numerical cognition, problem solving, the ability to reason, the ability to make decisions, meta-cognition, social cognition and language.
\end{LONG}
Among the details reported on the methodology, he mentions that incorrect responses typically produced auditory feedback, frequently accompanied by a time-out period, but that no other punitive method was used to promote productivity, accuracy or rapid responding. Lastly, he describes evidence that the nonhumans are not only motivated by food rewards, but also by the enjoyment of the tasks themselves: when given a choice between completing trials for pellets or receiving pellets for free but not being able to play the game-like tasks during the free-pellet period, the monkeys chose to work for their reward.
\begin{LONG}

The use of digital applications might benefit nonhuman in less directed ways too, by raising awareness and respect of t their cognitive abilities among the public. \citet{2021-EIT-ImprovintEthicalAttitudesToAnimalsWithDigitalTechnologiesTheCaseOfApesAndZoos-CoghlanWebberCarter} examined how digital technologies can be used to improve ethical attitudes towards nonhumans (focusing on nonhuman apes kept in zoos) by introducing digital technologies in zoos for both animal enrichment and visitor education.
\end{LONG}
\begin{BRIDGEPARAGRAPH}
Both analogical and digital setups must be careful to avoid experimental biases: we describe two particularly relevant ones to this work in the next section.
\end{BRIDGEPARAGRAPH}

\subsection{Experimental Biases}
\label{sec:experimentalBiases}

The history of Comparative Psychology has been prone with fights about the validity of methodologies and results:
\citet{2017-Psychonomic-AnimalLanguageStudiesWhatHappened-Pepperberg} describes various such tensions between researchers  about the cognition of animals, with some accusing other researchers in the field to be ``\emph{liars, cheats and frauds}'', and she highlights how sign language researchers were accused of ``\emph{cuing their apes by ostensive signals}'' and of ``\emph{consistently over-interpreting the animals' signs}''.
\begin{HEADPARAGRAPH}
We explore here two issues relevant to the experimentation protocol described in this work, namely \emph{selective reporting bias} (Section~\ref{sec:selectiveReporting}) and the \emph{``Clever Hans'' effect} (Section~\ref{sec:cleverHansEffect}).
\end{HEADPARAGRAPH}

\subsubsection{Selective Reporting Bias}
\label{sec:selectiveReporting}
Selection biases occur in a survey or experimental data when the selection of data points is not sufficiently random to draw a general conclusion.  Selective reporting biases are a specific form of selection bias whereby only interesting or relevant examples are cited.
Cognitive skills can be particularly hard to study in nonhumans, requiring unconventional approaches but often presenting the risk of such biases.
For example, an experimenter who would present a subject repeatedly with the same exercise could be tempted to omit or exclude bad performances (eventually attributing them to a ``bad mood'' of the subject, which stays a real possibility) and report only on good performances, creating a biased representation of the abilities of the subject, a selective reporting bias.
\begin{LONG}

Whereas \citet{2007-Methods-CreativeOrCreatedUsingAnecdotesToInvestigateAnimalCognition-BatesByrne} defends the use of anecdotes in comparative psychology, he does so ``\emph{provided certain conditions are met}'' so that to avoid such biases. 
defining an \emph{anecdotal method} in five steps:
\begin{enumerate}
\item Source Material Assembly;
\item Definition of the extraction process;
\item Categorization of extracted records;
\item Labeling of each record with a level of evidence (from ambiguous to highly suggestive);
\item Hypothesis generation to guide future studies.
\end{enumerate}
He emphasises the ``\emph{need to use neutral descriptions of behaviour that avoid implicit interpretation, recorded by experienced and knowledgeable observers \textbf{immediately after the occurrence of the event}.}'', and that all observations of rare events should be made available for later analysis by the anecdotal method.
\end{LONG}
We describe how to use a digital application to systematically log the result and easily avoid such bias in Section~\ref{sec:incaWhatIsMore}.
\begin{BRIDGEPARAGRAPH}
Another type of bias is that of the subject reading cues from the experimenter, which we describe in the next section.
\end{BRIDGEPARAGRAPH}

\subsubsection{``Clever Hans'' effect}
\label{sec:cleverHansEffect}

Among such methodological issues resulting in experimental biases, the most iconic one might be the eponymous horse nicknamed ``Clever Hans'' which appeared to be able to perform simple intellectual tasks, but in reality relied on involuntary cues given by not only by their human handler, but also a variety of human experimenters\begin{LONG}, as related by  \citet{2015-BiologicalTheory-Clever_Hans_Alex_the_Parrot_and_Kanzi_WhatCanExceptionalAnimalLearningTeachUsAboutHumanCognitiveEvolution-Trestman}:
\begin{quote}
``\emph{In the early years of the 20th century an unlikely and controversial celebrity rose to fame in Germany and internationally: a horse, aptly named Clever Hans, who apparently displayed a startling aptitude in mathematics as well as music theory, not to mention the ability to identify colors and individual people by name, read German, and answer a variety of questions in normal spoken German.  He responded to questions primarily via a code based on repeatedly tapping his right hoof, in combination with other responses such as nodding or shaking his head to indicate yes or no, and pointing to objects or photographs with his nose.}''
\end{quote}
The story is famous of course for how it illustrates how nonhumans can often more easily learn to read cues from the experimenter than to solve the problem asked from them\end{LONG}.
\begin{LONG}
One ignores such rules not only at its own risk, but at the risk of hurting a whole research area: in her recapitulation of the history of animal language studies, \citet{2017-Psychonomic-AnimalLanguageStudiesWhatHappened-Pepperberg} describes how coverage of issues about the validity of Comparative Psychology methodologies and results in the public media in 1980 moved government agencies to respond to the blow-back by cutting off the funding for all of the related studies.
\end{LONG}
\begin{LONG}
While issues such as over interpreting subject's signs or selectively reporting experimental reports can be avoided with the appropriate amount of rigor (eventually with some computerized help, as discussed in Section~\ref{sec:automaticLog}), avoiding having subjects reading experimenter's cues requires special care to be taken when designing the experimental protocol: we describe in the next section some guiding principles which exclude the very possibility of such biases from experimentation results.
\subsection{Masked Experimental Protocols}
\label{maskedExperimentalProtocols}
\end{LONG}
It is possible to avoid the confusion between a subject's ability to read cues from the experimenter from its ability to answer the tests presented to them by such an experimenter. The principle is quite simple: make sure that the experimenter does not know the test, by having a third party out of reach from the subject's reading to prepare the test.
Whereas such experimental setup was historically referred to as a "Blind Setup" or a "Blinded Setup", we follow the recommendations of Moris et al.~\cite{2007-MaskingIsBetterThanBlinding-MorrisWormald} and prefer the term of "masked" to the term "blind" when describing the temporary and purposeful restricted access of the experimenter to the testing information.
\begin{LONG}

In an analogical setting, the set-up of a masked experimental protocol is more costly than that of less careful ones. For example, \citet{2018-ACI-AdvancingCommunicationWithBirdsCanTheyLearnToRead-CunhaClubb} describe an experimental protocol where an assistant prepares a pile of prompt cards, which the experimenter presents to the subject without looking at their content until after the subject responded, in order to know whether to praise and reward them or not.
We describe our digital set-up for a masked experimental protocol in Section~\ref{sec:maskedExperimentalSetup}: the digital device completely replaces the assistant, and assists the experimenters, telling him whether the subject answered correctly or not. As long as the device is placed in such a position that the subject can see the visual display but the experimenter cannot, there is physically no way for the subject to read cues from the experimenter, hence avoiding the \emph{``Clever Hans'' effect}.

\end{LONG}
In the next section, we describe an application designed so that to facilitate a type of ``masked'' experimental set-up, in which it is guaranteed that the ability of the subject to read cues from the experimenter does not affect the result of the experiment, as the experimenter himself ignores the question (and hence its correct answer) being asked to the subject.

\section{Application}
\label{sec:incaWhatIsMore}

\begin{HEADPARAGRAPH}
We developed a web application
\begin{NONANONYMOUS}
\texttt{InCA-WhatIsMore}
\end{NONANONYMOUS}
as a simple combination of \texttt{JavaScript}, \texttt{CSS} and \texttt{HTML} using libraries from the \texttt{Svelte} project, made available on a simple web-page. While its simple structure (described in Section~\ref{sec:applicationStructure}) was originally developed as a simple mock-up to visualize how a simple web application could help setting up masked experiments (described in Section~\ref{sec:maskedExperimentalSetup}) with extensive logging abilities (described in Section~\ref{sec:loggingStructure}), it was found complete enough to be used as a final application, and the structure to be simple enough that even the subjects themselves could navigate it.
\end{HEADPARAGRAPH}

\subsection{Application's Structure}
\label{sec:applicationStructure}

\begin{figure}
\begin{minipage}[b]{.5\linewidth}
\begin{minipage}[b]{1\linewidth}
\begin{NONANONYMOUS}%
\includegraphics[width=\textwidth]{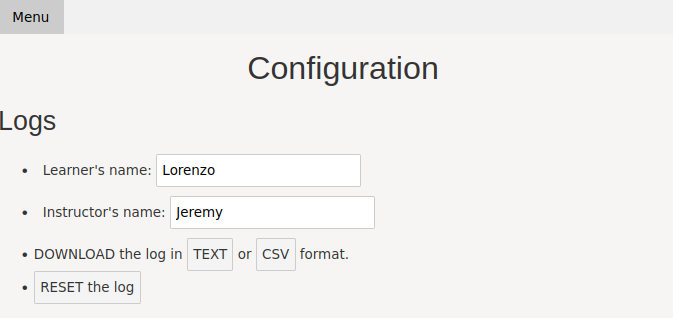}
    \Description{Screenshot of the first third of the settings menu.}
\caption{The logs are exported in the top part of the setting page of the application.\label{fig:ScreenshotOfLogGeneration}}
\end{NONANONYMOUS}\begin{ANONYMOUS}%
    \Description{Screenshot of the first third of the settings menu, dedicated to the log generation.}
\includegraphics[width=\textwidth]{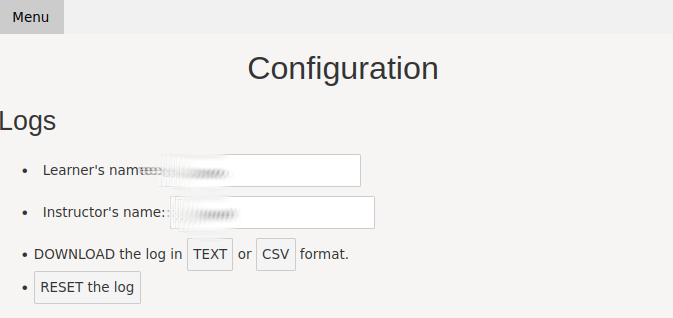}
\caption{The logs are exported in the top part of the setting page of the application. The names were blurred to protect the anonymity of the submission.\label{fig:ScreenshotOfLogGeneration}}
\end{ANONYMOUS}%
\end{minipage}\\
\begin{minipage}[b]{1\linewidth}
\includegraphics[width=\textwidth]{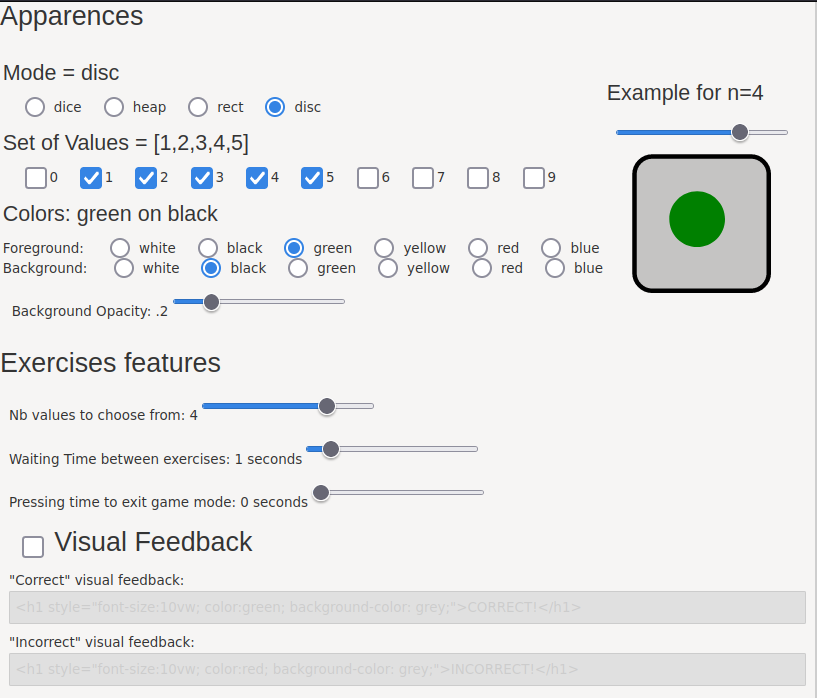}
    \Description{Screenshot of the second third of the settings menu, dedicated to the appearance and difficulty of exercises.}
\caption{The part of the setting page dedicated to the appearance and difficulty of the exercises.\label{fig:AparencesExercisesFeatures}}
\end{minipage}
\end{minipage}
\hfill
\begin{minipage}[b]{.45\linewidth}
\includegraphics[width=\textwidth]{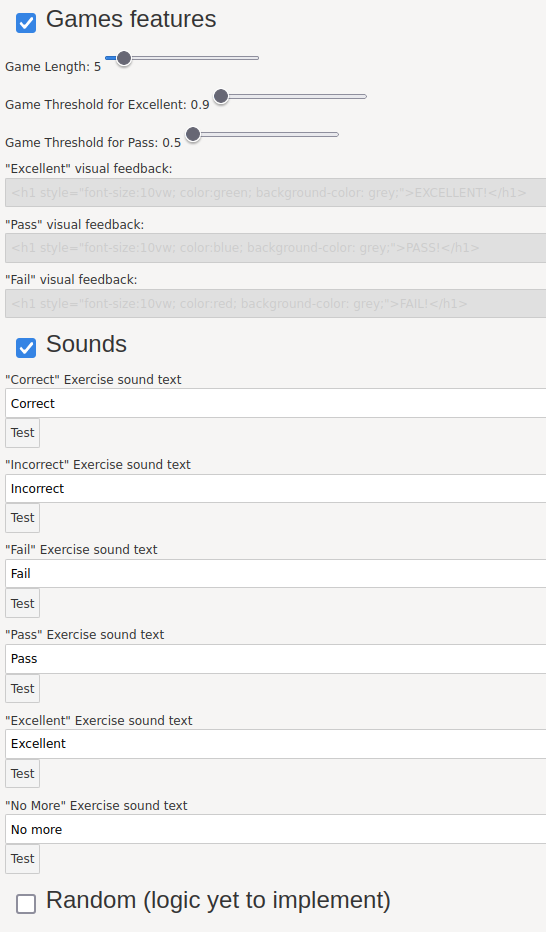}
    \Description{Screenshot of the last third of the settings menu, dedicated to game features and sound feedback.}
\caption{The part of the setting page dedicated to the game features and sound feedback.\label{fig:GamesSoundRandom}}
\end{minipage}
\end{figure}

\begin{figure}
\begin{minipage}[b]{.3\textwidth}
\centering
\includegraphics[width=\textwidth]{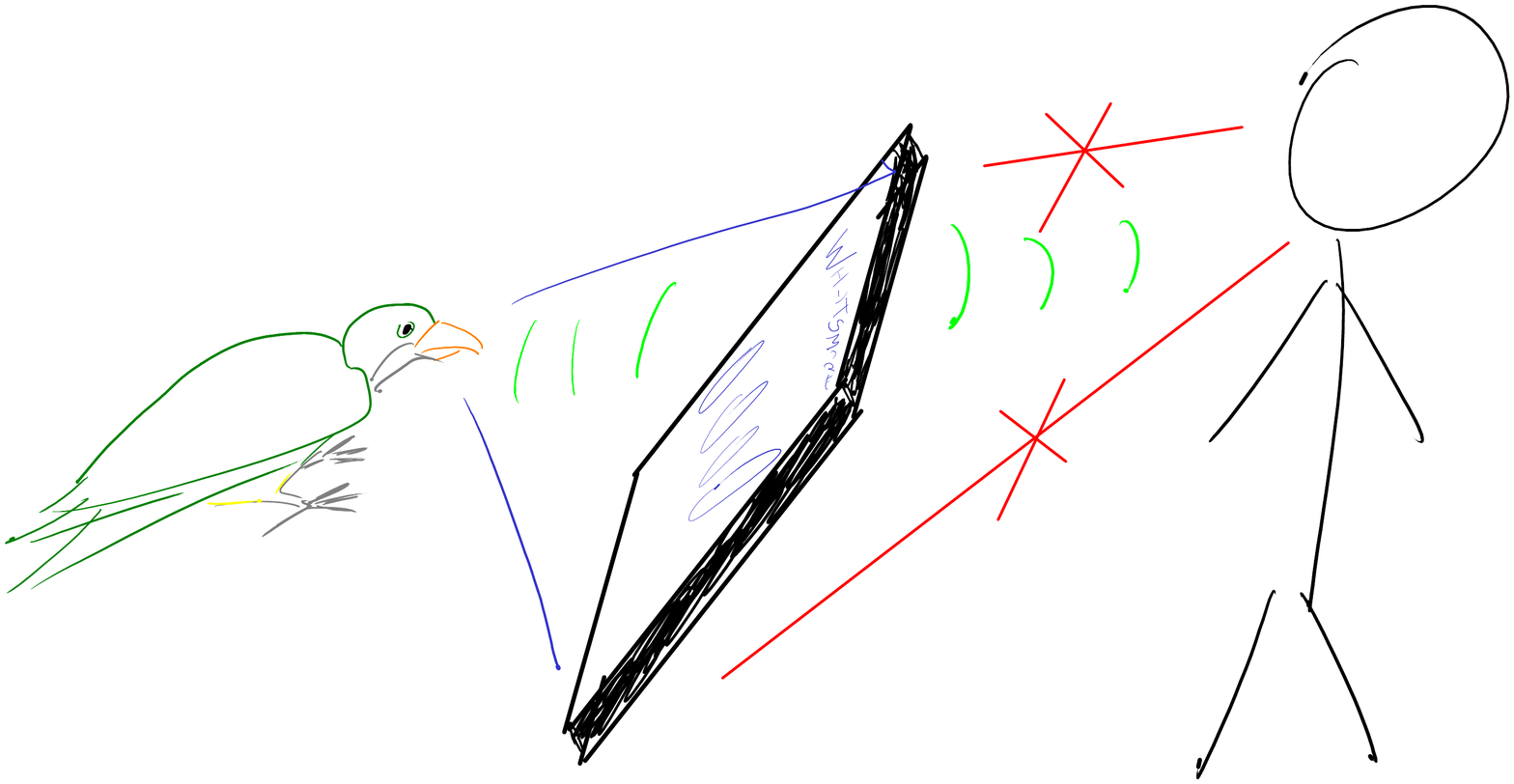}
\Description{Artistic Rendition of a Masked Experimental Setup.}
\caption[Artistic Rendition of a Masked Experimental Setup]{The Masked Setup. The subject (left) can see the display and hear  the device (center), but the experimenter (right) can only hear  the device and not see its display.\label{fig:maskedExperimentalSetupDrawing}}
\end{minipage}
\hfill
\begin{minipage}[b]{.3\textwidth}
\centering
  \includegraphics[width=\textwidth]{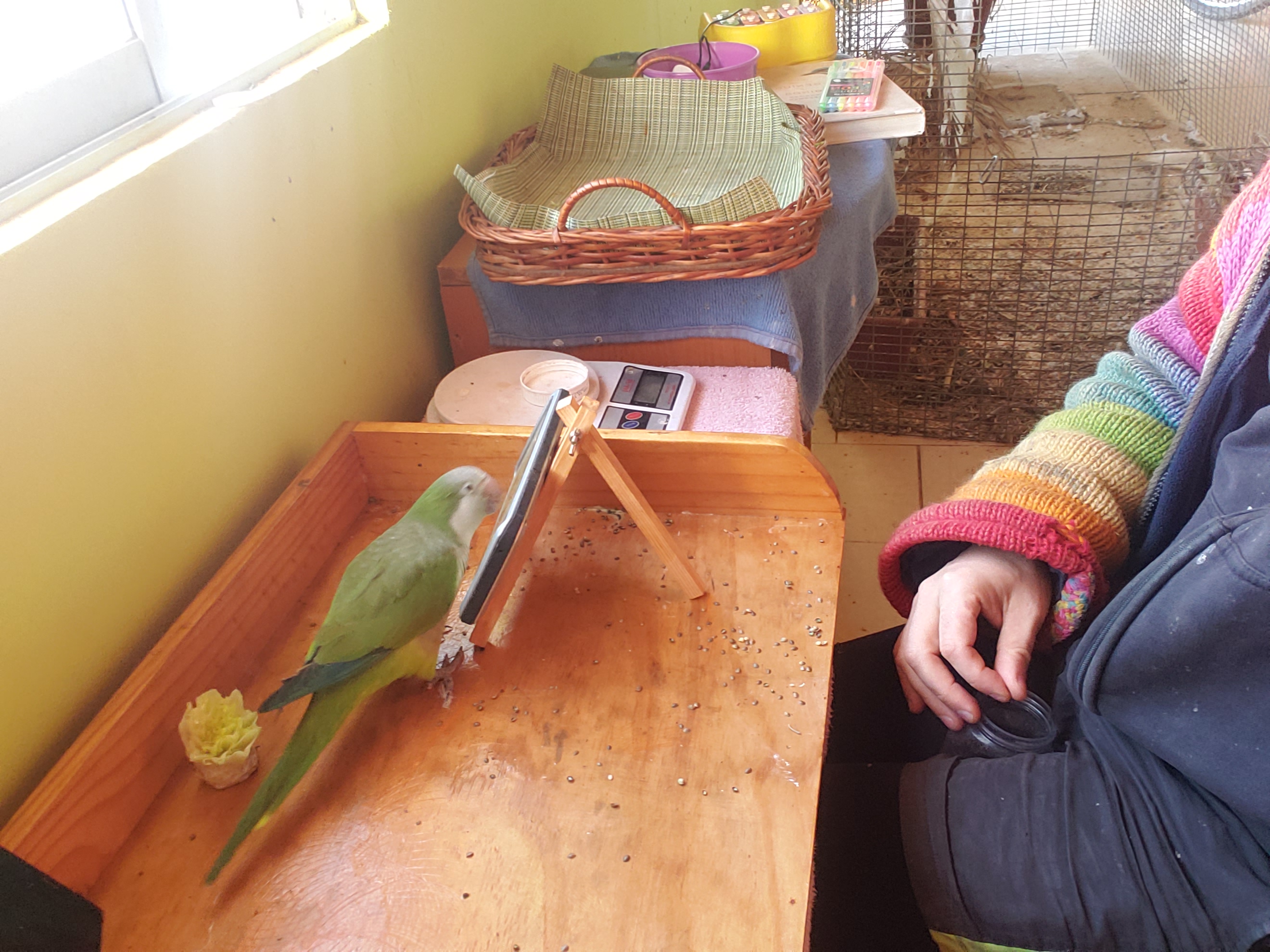}
  \Description{Picture of a Masked Experimental Setup with one Monk Parrot, one digital device, and one experimenter who cannot see the screen of the device.}
    \caption[Example of Masked Experimental Setup]{Example of masked experimental set-up: the experimenter can hear the instructions from the device and encourage the subject, but cannot give any cue about  correct answers.\label{fig:maskedExperimentalSetupPicture}}
\end{minipage}
\hfill
\begin{minipage}[b]{.3\textwidth}
\centering
\includegraphics[width=\textwidth]{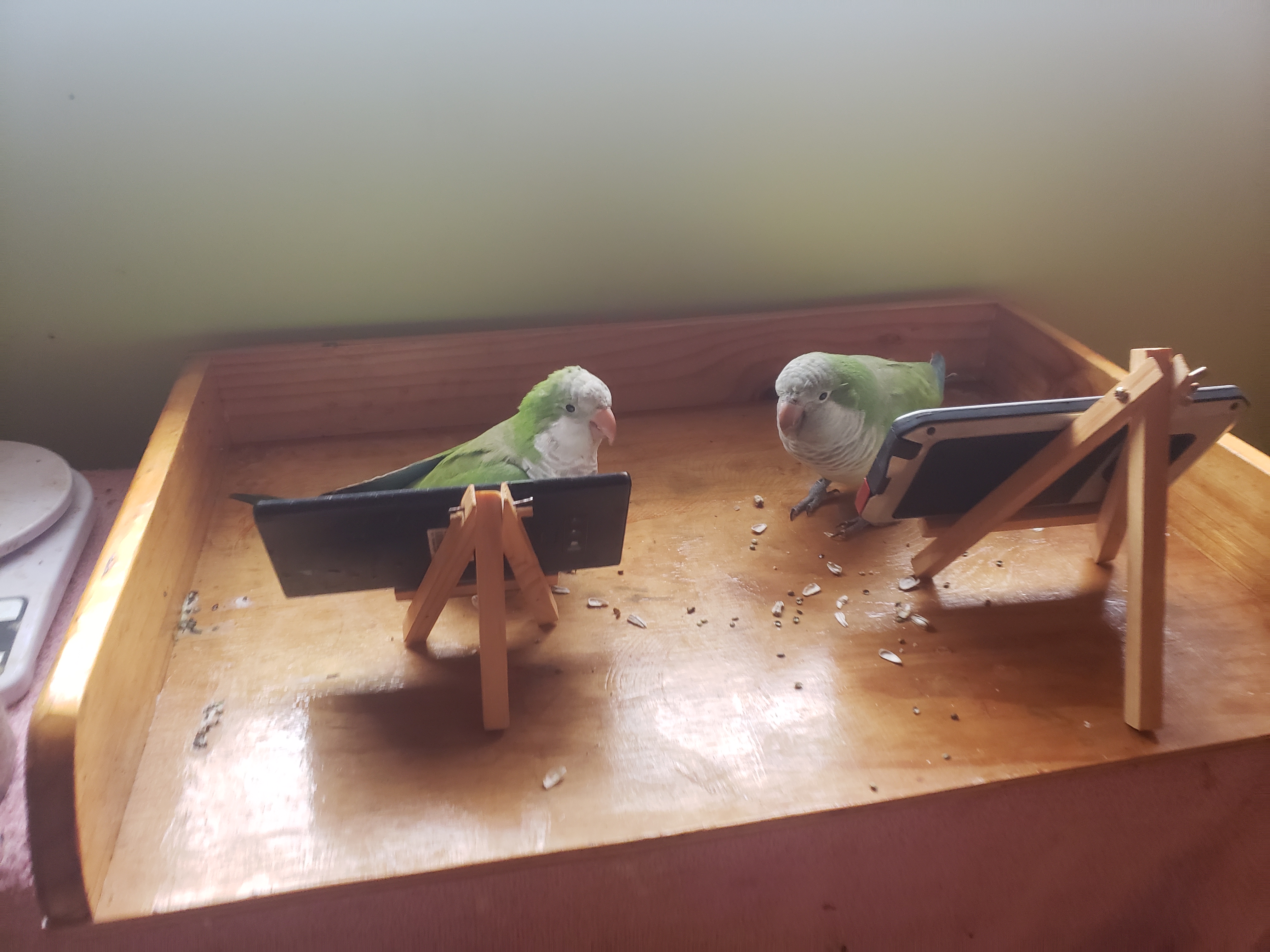}
  \Description{Picture of the experimenter's view in  a Masked Experimental Setup with two Monk Parrots and two digital devices.}
\caption[Experimenter View of  Masked Experimental Setup]{The  masked experimental set-up as viewed by the experimenter, with two subjects participating in the experiment at the same time, each with its own device.\label{fig:experimenterViewOfTheMaskedExperimentalSetup}}
\end{minipage}
\end{figure}

The web application is composed of four views. The first two, the \texttt{Main Menu} (described in Figures~\ref{fig:screenshotOfStartingMenu} and~\ref{fig:AparencesExercisesFeatures}) and the \texttt{Gaming View} (which can be seen in Figures~\ref{fig:lorenzoSelectingLargestValueOutOfFourDiscs} and~\ref{fig:maskedExperimentalSetupPicture} among others), are especially designed to be navigable by nonhuman subjects. The access to the two others, the \texttt{settings} (see Figures~\ref{fig:ScreenshotOfLogGeneration} to~\ref{fig:GamesSoundRandom}) and the \texttt{information} views are tentatively restricted to the experimenters by requesting the long pressing of a button.

\begin{figure}
    \centering
    \begin{minipage}[b]{.45\linewidth}
    \begin{NONANONYMOUS}
    \includegraphics[width=\textwidth]{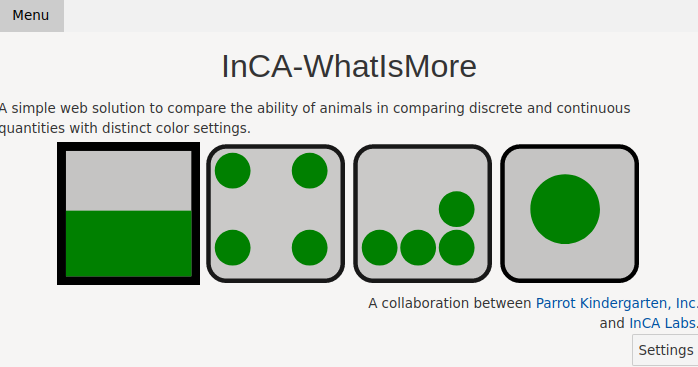}
    \Description{Screenshot of the menu of the application ``What is more''.}
      \caption[Nonhuman friendly menu]{The main menu of the application is designed so that the subject can choose in which visualisation mode it wishes to play, in the hope to support a sense of agency.  \label{fig:screenshotOfStartingMenu} }
    \end{NONANONYMOUS}
    \begin{ANONYMOUS}
    \includegraphics[width=\textwidth]{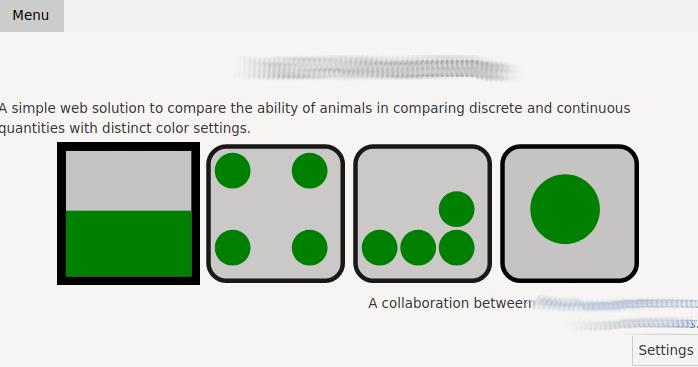}
    \Description{Screenshot of the menu of the web application.}
    \caption[Nonhuman friendly menu]{The main menu of the application is designed so that the subject can choose in which visualisation mode it wishes to play, in the hope to support a sense of agency. The name of the application and the collaborations were blurred to protect the anonymity of the submission.  \label{fig:screenshotOfStartingMenu} }
    \end{ANONYMOUS}
    \end{minipage}\hfill
\begin{minipage}[b]{.45\textwidth}
\centering
\begin{NONANONYMOUS}
  \includegraphics[width=\textwidth]{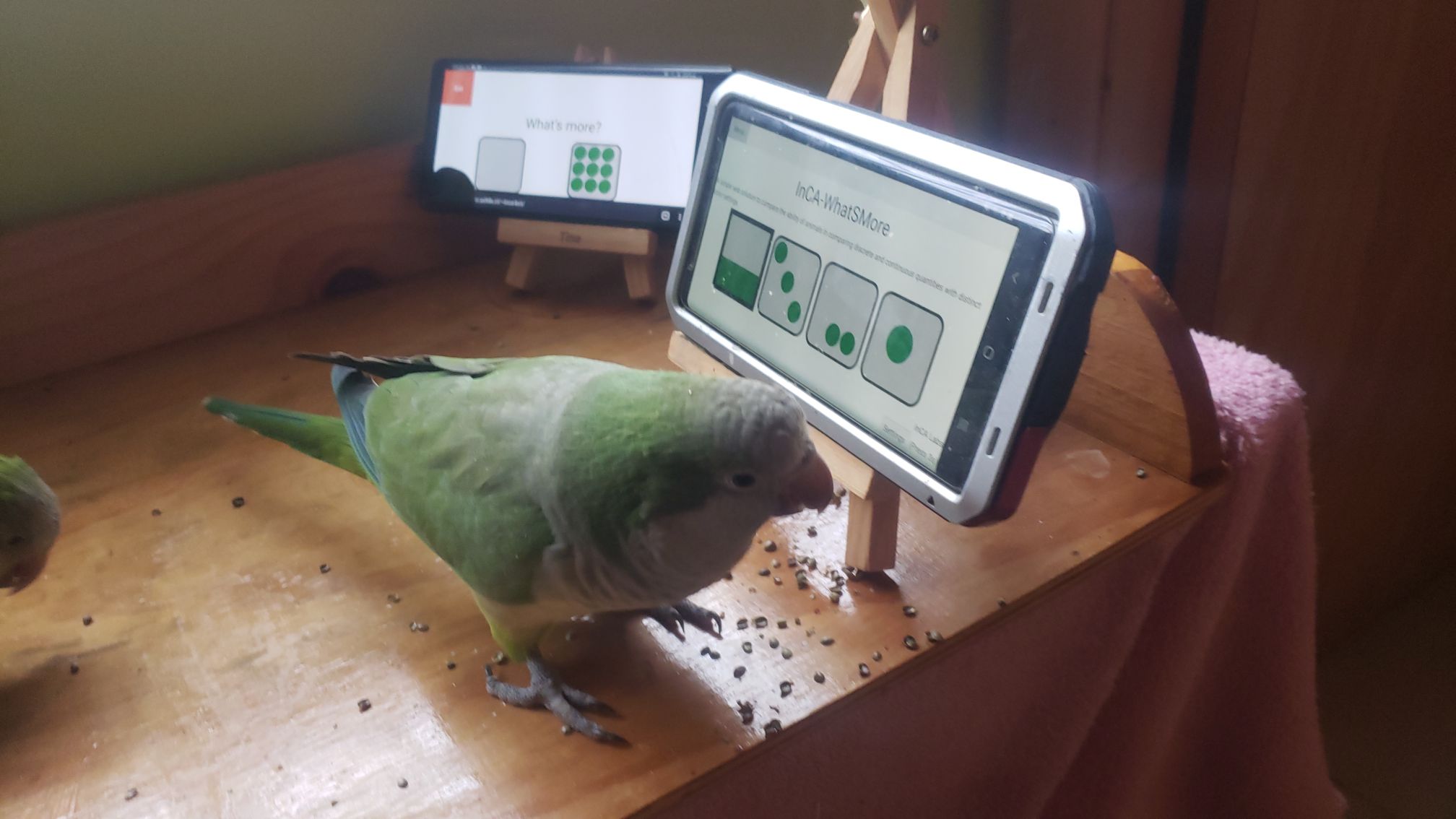}
    \Description{Monk Parakeet choosing an option in the menu of the application ``What is More''.}
\caption[Subject using the menu to choose display mode]{Both subjects quickly learned to select a display mode to start a game, but did not seem to show a preference for a display mode in particular.\label{fig:LorenzoChoosingModeInWhatIsMore}}
\end{NONANONYMOUS}
\begin{ANONYMOUS}
  \includegraphics[width=\textwidth]{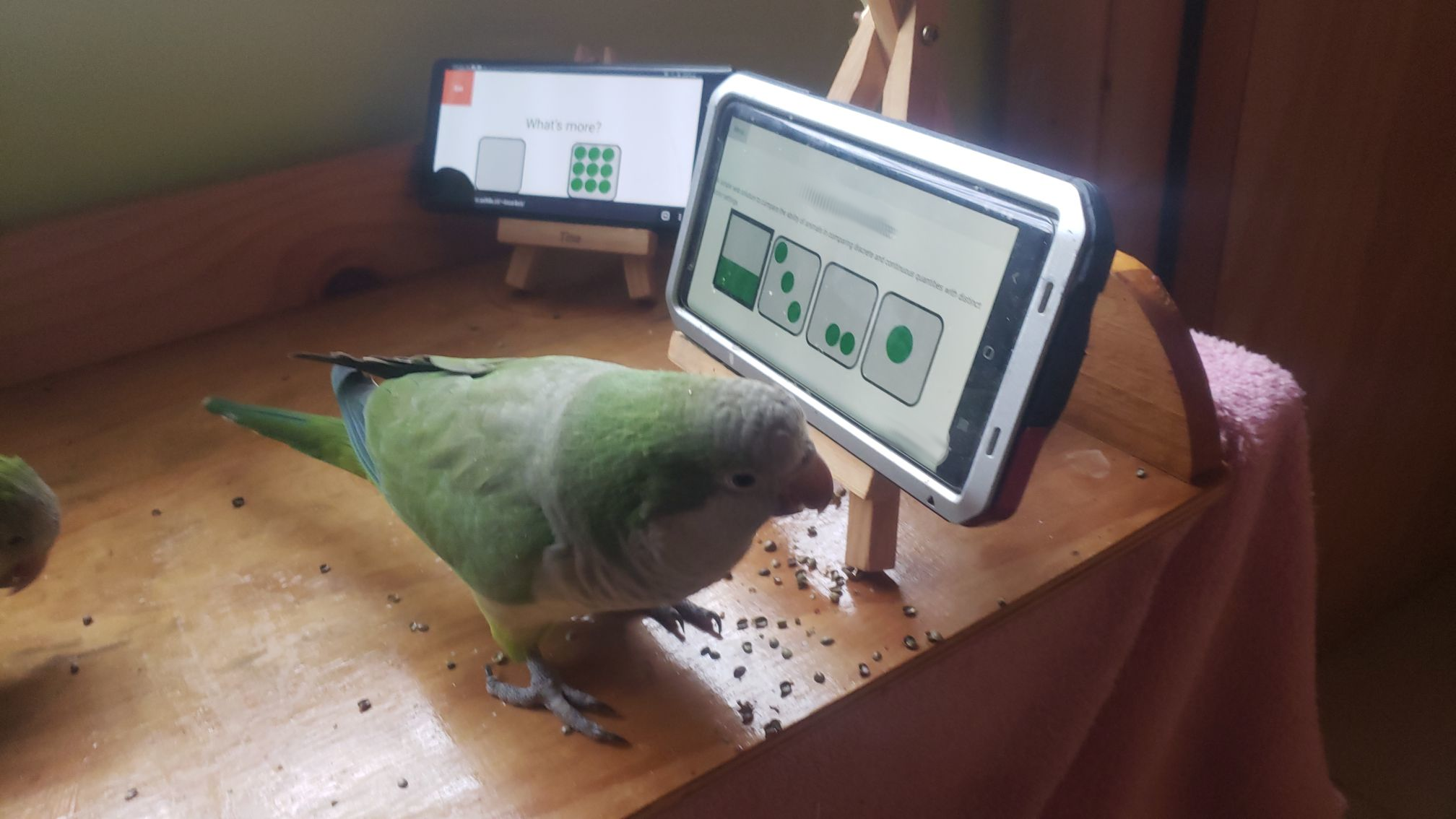}
    \Description{Monk Parakeet choosing an option in the menu of the application.}
\caption[Subject using the menu to choose display mode]{Both subjects quickly learned to select a display mode to start a game, but did not seem to show a preference for a display mode in particular. The name of the application and the collaborations were blurred to protect the anonymity of the submission.\label{fig:LorenzoChoosingModeInWhatIsMore}}
\end{ANONYMOUS}
\end{minipage}
    \end{figure}

\begin{figure}
\begin{minipage}[b]{.45\linewidth}
\includegraphics[width=\textwidth]{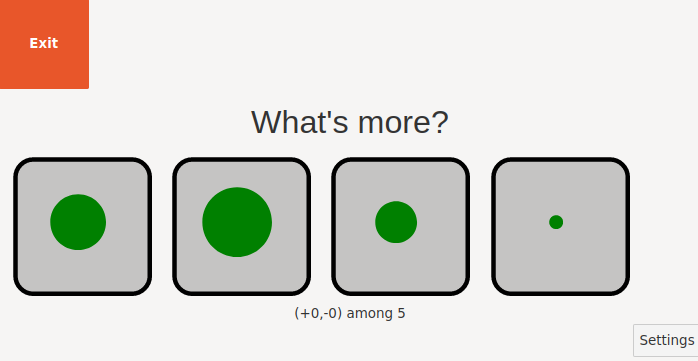}
\Description{Screenshot of the game view of the application.}
  \caption{A screenshot of the game view of the application, asking to choose the largest disk out of four.
  Top left is the orange ``Exit'' button actionable by the subject.
  Bottom right is the setting button requesting a long pressure to be activated.
Bottom center is a summary of the game score.\label{fig:screenshotGameViewFourDisk}}
\end{minipage}\hfill\begin{minipage}[b]{.45\linewidth}
\includegraphics[width=\textwidth]{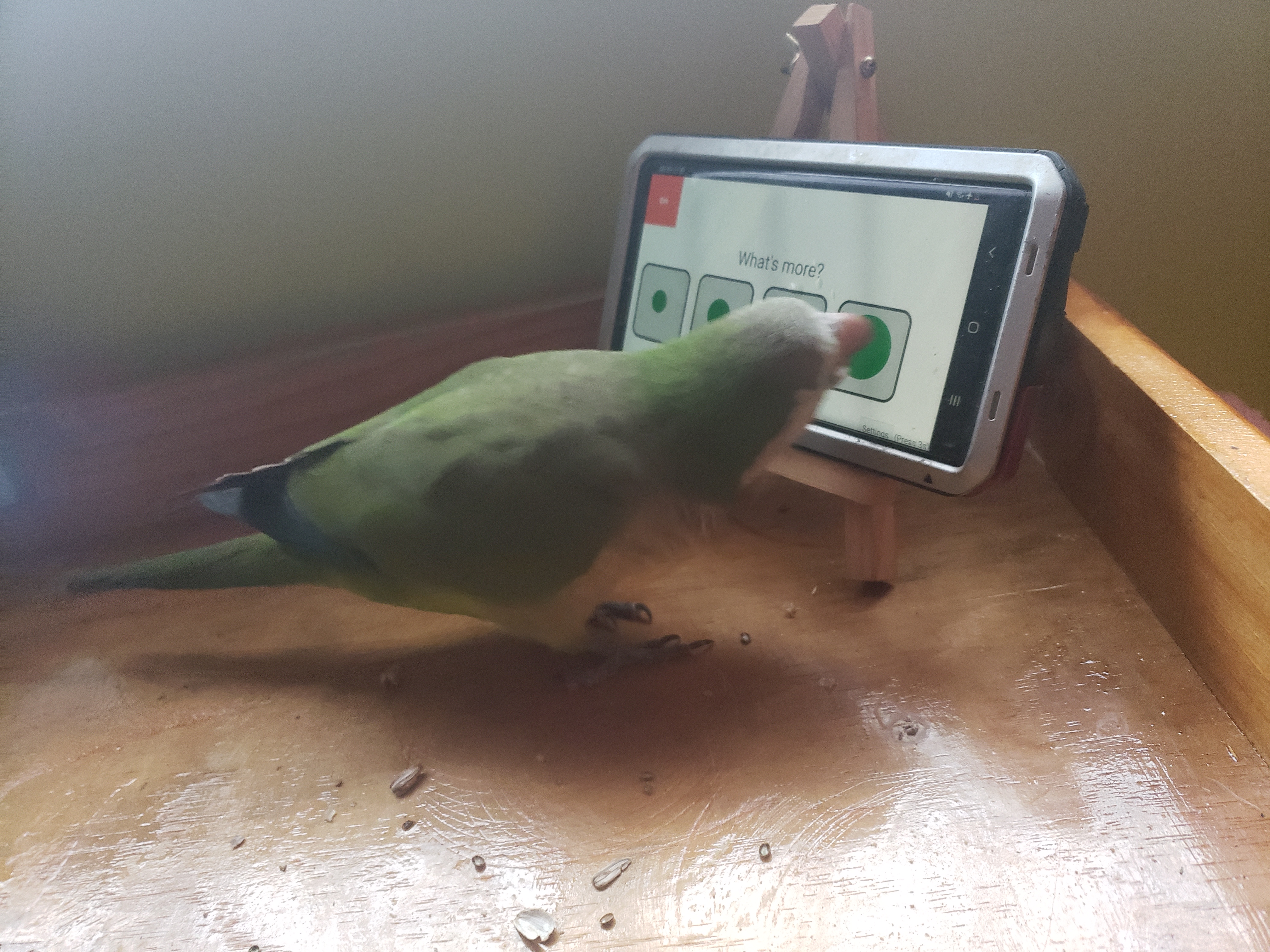}
\Description{Picture of a Monk Parakeet interacting with the  game view of the application, with the display mode ``Disc''.}
\begin{ANONYMOUS}
\caption{Monk Parakeet selecting the largest disc out of four.\label{fig:lorenzoSelectingLargestValueOutOfFourDiscs}}
\end{ANONYMOUS}\begin{NONANONYMOUS}
\caption{The Monk Parakeet Lorenzo selecting the largest disc out of four. \label{fig:lorenzoSelectingLargestValueOutOfFourDiscs}}
\end{NONANONYMOUS}
\end{minipage}
\end{figure}

\begin{LONG}
\subsubsection{Main Menu}
\label{sec:mainMenu}

The view of the \texttt{main menu} is accessed when the application is opened: see Figure~\ref{fig:screenshotOfStartingMenu} for a screenshot, and Figure~\ref{fig:LorenzoChoosingModeInWhatIsMore} for a picture of a subject using it to select a display mode. From this view, the user can navigate to the other views of the application. On the center of the screen  are four figures, each one representing a different visualisation mode used on a random value.
Two of such modes are discrete: one is representing the value as a number of dots on a dice face, the other on a 3 by 3 grid (as that of the dice) but as a heap of dots. The two other modes are more continuous: one is representing each value by a recipient more or less filled with liquid according to the value, the other by a circle of radius growing with the value.
The user (let it be the experimenter or the subject) can pick any of the four display modes in order to start the game in it.  At the bottom of the screen stands a button to access the settings section, activated after a long press (which length is set up in the setting view).

\subsubsection{Gaming view}
\label{sec:playingScreen}

The most important view is the \texttt{gaming} view, allowing the subject to ``play'': see Figure~\ref{fig:screenshotGameViewFourDisk} for a screenshot, and Figures~\ref{fig:teaser2}, \ref{fig:teaser3}, \ref{fig:plankSetup}, \ref{fig:stableSetup}, and \ref{fig:lorenzoSelectingLargestValueOutOfFourDiscs} for pictures of subjects playing the game.
The view displays a set of values in some display mode, requesting the user to choose the largest one. Each action triggers an audio feedback, indicating if it was correct or wrong. After a given number of exercises, the game ends and give an audio feedback about how the score for this game placed with two boundaries (boundaries which can be modified in the settings page, as well as the words being vocalized in each audio feedback).
The view has also an exit button on the top left corner intended to be usable by the user, and a settings button actionable by long pressing it by a parameterized amount of time.

\subsubsection{Settings}
\label{sec:settings}

Whereas the subject can choose the display mode in which they prefer to play, and exit the game view to change it at any time, other settings are accessible on a more technical view, designed for the experimenter to set-up other aspects of the software: the \texttt{settings} view (Figures~\ref{fig:AparencesExercisesFeatures} and~\ref{fig:GamesSoundRandom}).
As the visual and sound outputs of touch screen devices were designed for humans, and as very little is known about the adequateness of such outputs for nonhumans, the software was designed so that to maximize the control given to each experimenter on the visual and sound output of the application, so that each experimenter can find the most adapted settings to the characteristics of the species and/or subjects with which the software is used.
As such, the software permits the experimenter to change, among other things, the color schemes and the number of the values displayed, the domain from which such values are randomly chosen, and the number of questions before a game is ended: it is hoped that such parameters will be useful in future studies.

\subsubsection{About}
\label{sec:InfoView}

This last view, a priory accessed only by the experimenters, displays various information about the application, such as its version, an overview of features soon to be added and recently added, instructions of usage use, references and acknowledgments to collaborators.

\end{LONG}

\begin{BRIDGEPARAGRAPH}
We describe how to use such a web application so that to implement a masked experimentation protocol where the subject cannot read any clue from the experimenter, because the experimenter ignores the question given to the subject.
\end{BRIDGEPARAGRAPH}

\subsection{Masked Experimental Setup}
\label{sec:maskedExperimentalSetup}

Among other features, the web application was designed to facilitate digital experiments similar to that performed by \citet{2008-AC-TheDiscriminationOfDiscreteAndContinuousAmountsInAfricanGreyParrots-AlAinGiretGrandKreutzerBovet} but in a way such that the experimenter does \emph{not} know what side the ``correct'' answer is, a masked experimental setup. This insures that the subject cannot receive any voluntary or involuntary cue from the experimenter.
Such a purpose is achieved through the extensive audio feedback system, which aims at notifying the experimenter about any event which requires their intervention (e.g. rewarding or encouraging the subject, or acknowledging that the subject does not want to play this game any more), so that they do not need to check the screen of the device at any point.

\subsection{Logging structure}
\label{sec:loggingStructure}

In non digital experiments in comparative psychology, the experiments is usually recorded on video so that the video recording can be later processed in order to generate an extensive log of the interactions of the subject during the experiment. Such a task is long and tedious, and no video processing software is yet able to automatize such a process. An important advantage of a digital experimental set-up such as that allowed by the software \begin{NONANONYMOUS}\texttt{InCA-WhatIsMore}\end{NONANONYMOUS} is the ability to \emph{automatically} log the interactions of the subject with the application.
\begin{LONG}

The logs are exported in the top part of the setting page of the application (previously described in Figure~\ref{fig:ScreenshotOfLogGeneration}, on page~\pageref{fig:ScreenshotOfLogGeneration}).
Two formats are available:
  the first format, \texttt{.txt} is designed to be easily readable by humans; while 
  the second format, \texttt{.csv} is more adequate for machine processing.
  \end{LONG}
%
%
The software \begin{NONANONYMOUS}\texttt{InCA-WhatIsMore}\end{NONANONYMOUS} generates logs with data to be analyzed by researchers, including information on both the test performed and the subject's performance (see Figure~\ref{fig:shortExampleOfLog} for a short extract\begin{LONG}, and Figure~\ref{fig:pageLongLogExample} for a longer one\end{LONG}).

\begin{figure}
\centering
\begin{adjustbox}{max width=\linewidth}
\begin{BVerbatim}
Test no, Test Name, Learner, Trainer, C_0, C_1, C_2, C_3, C_4, Value selected , Correction , Date, Answering Time (ms), Other Parameters
1, dice, Subject, Experimenter, 1,4,,,, 4,true, [2022-05-19 17:02(25.981)], 7946, background black, foreground green, bg opacity .2, Value Set [1,2,3,4,5]
(...)
81, rect, Subject, Experimenter, 4,2,3,,, 3,false, [2022-05-19 17:26(55.124)], 4655, background black, foreground green, bg opacity .2, Value Set [1,2,3,4,5]
(...)
180, heap, Subject, Experimenter, 3,2,1,,, 2,false, [2022-05-19 17:35(06.6)], 926, background black, foreground green, bg opacity .2, Value Set [1,2,3,4,5]
\end{BVerbatim}
\end{adjustbox}
\caption{A short extract showing four selected lines of the log generated by the application for the afternoon session of the 19th of May 2022 (deleted blocks of lines are marked by ``\texttt{(...)}''). See Figure~\ref{fig:shortExampleOfLog} for a more readable reformatting of the same extract. Log entries such as ``\texttt{background black, foreground green, bg opacity .2}'' refer to visualisation options, not used in this work. \label{fig:shortLogExampleVerbatim} }
\end{figure}

\begin{figure}
\resizebox{\columnwidth}{!}{%
\tiny
  \begin{tabular}{*{20}{c}}
Test no & Test Name & C0 & C1 & C2 & C3 & C4 & Value selected & Correction & Date                       & Other Parameters      \\ \hline
1       & dice      & 1  & 4  &    &    &    & 4              & true       & [2022-05-19 17:02(25.981)] & Value Set [1,2,3,4,5] \\
81      & rect      & 4  & 2  & 3  &    &    & 3              & false      & [2022-05-19 17:26(55.124)] & Value Set [1,2,3,4,5] \\
180     & heap      & 3  & 2  & 1  &    &    & 2              & false      & [2022-05-19 17:35(06.6)]   & Value Set [1,2,3,4,5] \\
\end{tabular}%
}
\caption{A more readable format of the log extract from Figure~\ref{fig:shortLogExampleVerbatim}, with less relevant columns removed for readability.  
  Observe that the subject was offered to choose the largest between 2 (on the first test) and 3 (on the $81$st and $180$th tests) values, represented as \texttt{dice} (first test), \texttt{rect} ($81$st test) and \texttt{heap} ($180$th test), and that the subject chose once correctly, and two times incorrectly, in games where the values were taken from the set $\{1,2,3,4,5\}$, with the precise time and date of each answer duly recorded. The columns labeled C3 and C4 are empty because no test was performed requesting the subject to choose the maximal value between 4 or 5.
  \label{fig:shortExampleOfLog}}
\end{figure}

\begin{LONG}
We describe here the format of the logs in versions $2.x$ of the software.
The first three rows generated within the log indicate information about the test:
\begin{itemize}
    \item \textbf{Test Name:} The test performed, the value is the type of representation, this can be "dice", "heap", "rect" or "disc".
    \item \textbf{Learner:} The name of the test subject, used to subsequently run analyses such as performance over time or to differentiate its statistics.
    \item \textbf{Trainer:} The name of the experimenter. This could be used in later studies where various experimenters apply the same test to the same subject, to check for variance in performance from one experimenter to the other.
      \end{itemize}
      
The following columns indicate quantitative information about the distribution of quantities within the test, as well as information about the test subject's performance.
\begin{itemize}
    \item \textbf{C0,C1,C2,C3:} The qualitative representation of the quantities delivered, these can be discrete or their representation in continuous quantities, the order of distribution of these quantities also indicates the order deployed within the application, the values being ordered from left to right.
    \item \textbf{Value Selected: } The value chosen by the test subject.
    \item \textbf{Correction: } The correctness of the value selected by the test subject, being "true" if it is the largest amount and "false" otherwise.
\end{itemize}

The last columns indicate qualitative values about the test, these values provide information about both the performance and the setup of the test.
\begin{itemize}
    \item \textbf{Date:} The date on which the test was performed, in timestamp format, including  the precise time in milliseconds.
    \item \textbf{Answering Time (ms):} The time it takes for the test subject to respond from the display of the quantities to be chosen, represented in milliseconds. Note that this is more precise than the simple difference between the times of two consecutive time stamps, as the application includes (parameterizable) waiting time between tests, a pause between games, break to return to the menu, etc.
    \item \textbf{Other Parameters:} Parameters that visually describe the display of the quantities, such as the color of the background and the color in which the representations of the quantities are displayed. These parameters were modified only in the development phase in order to find an adequate color scheme for the two subjects in particular (see Section~\ref{sec:sensorialDiversity} for a discussion of the sensory variability of general test subjects), but could be used in the future to adapt the software to other individuals, potentially from other species with distinct sensory ranges; and to formally study (ethically, see Section~\ref{sec:EthicalMeasurementOfUNabilities} for a discussion of the related challenges) the limits of the sensory range of any subject.
\end{itemize}
\end{LONG}

\begin{BRIDGEPARAGRAPH}
In the next section, we describe the training and experimental protocol which was used to generate the data measured by such logs.
\end{BRIDGEPARAGRAPH}
 \section{Experimentation Protocol}
\label{sec:experimentations}

\begin{HEADPARAGRAPH}
The experimental protocol was divided in three phases, which we describe in Section~\ref{sec:phases}.  The precautions taken to protect the well-being of the nonhuman subjects (described in Section~\ref{sec:ethicalPrecautions}) were validated by the Institutional Animal Care and Use Committee (IACUC) \begin{NONANONYMOUS} (in Spanish ``COMITÉ INSTITUCIONAL de CUIDADO y USO de ANIMALES (CICUA)'') \end{NONANONYMOUS} of the researcher's institution.  The statistical analysis (described in Section~\ref{sec:statisticalAnalysis}) were scheduled as part of the experimental protocol, independently from the results of the experiments.
\end{HEADPARAGRAPH}

\subsection{Phases of the protocol}
\label{sec:phases}

The protocol was implemented in three phases:
a phase of \emph{development} (of the software) with only one subject (the first one),
a phase of \emph{training} with two subjects and a mix of unmasked and masked protocols, and
a phase of \emph{testing} using the masked protocol and collecting data with both subjects.
\begin{LONG}
\begin{itemize}
\item During the \emph{development phase}, a single subject (hereafter referred to as ``the first subject'') interacted with the various prototypes of the software, in a non masked experimental setting where the experimenter could observe the display of the screen.  Each time the software was modified, it was tested by two humans subjects before being used by any of the nonhuman subjects, in order to minimize the potential frustration of the nonhuman subjects while using a malfunctioning application.
\item During the \emph{training phase}, both subjects were invited to use the software, each on its own device, in a non masked experimental setting where the experimenter could observe the display of the screen: see Figures~\ref{fig:teaser2}, \ref{fig:teaser3}, \ref{fig:plankSetup}, \ref{fig:stableSetup}, and \ref{fig:lorenzoSelectingLargestValueOutOfFourDiscs}  for pictures  of the set-up during the training phase.
\item During the \emph{testing phase}, both subjects were invited to use the software, each on its own device, this time in a masked experimental setting where the experimenter could not observe the display of the screen, so that they ignored the question asked to each subject and could not cue them, and limiting themselves to encourage and reward each subject according to the feedback vocalized by the application (see Figures~\ref{fig:maskedExperimentalSetupDrawing} to~\ref{fig:experimenterViewOfTheMaskedExperimentalSetup}, on page~\pageref{fig:maskedExperimentalSetupPicture}, for examples of masked experimental setups).
\end{itemize}
\end{LONG}
\begin{BRIDGEPARAGRAPH}
The subjects' welfare was cared for during each of those three phases: we describe some of the precautions taken in the next section.
\end{BRIDGEPARAGRAPH}

\subsection{Ethical Precautions}
\label{sec:ethicalPrecautions}

\begin{HEADPARAGRAPH}
Various precautions were taken to protect both the physical (Section~\ref{sec:physical-settings}) and psychological well-being (Sections~\ref{sec:appl-usab}, \ref{sec:sense-agency} and \ref{sec:appr-exper-prot}) of the subjects during the three phases of the project.
\end{HEADPARAGRAPH}

\begin{MAYB}
develop how this is in accordance both with the recommendations from Mancini et al.\cite{2022-FVS-TheCaseForAnimalPrivacyInTheDesignOfTechnologicallySupportedEnvironments-PaciManciniNuseibeh,2022-FAS-RelevanceImpartialityWelfareAndConsentPrinciplesOfAnAnimalCenteredResearchEthics-ManciniNannoni,2019-ACI-AMethodForEvaluatingAnimalUsability-RugeMancini,2017-TowardsAnAnimalCenteredEthicsForACI,2016-ACMI-IntroductionFrameworksForACIAnimalsAsStakeholdersInTheDesignProcess-NorthMancini} and the ACI board,
and in the interest of the validity of the experience and the mental health of the researchers.
\end{MAYB}

\begin{MAYB}
Mention \citet{2021-Area-BeyondHumanEthics-Oliver}
\cite{2019-ACI-AMethodForEvaluatingAnimalUsability-RugeMancini}
\cite{2017-TowardsAnAnimalCenteredEthicsForACI}
\end{MAYB}

\subsubsection{Physical settings}\label{sec:physical-settings}

\begin{LONG}
The subjects were hosted in a private residence counting with three aviaries, each large enough to allow some amount of flight:
  one ``laboratory'' building  with meshed windows  containing a ``Nest'' aviary with a simple door, of size $3\times1 \mbox{m}^2$ and $2\mbox{m}$ high, containing a nest, a plant and various toys and nesting material;
  one ``South'' aviary, corridor shaped with two sets of double doors, of $6\times 1 \mbox{m}^2$ of surface size and $2\mbox{m}$ high; and
  one ``North'' aviary with one set of double doors, of $6\times 3 \mbox{m}^2$ of surface size and $1\mbox{m}$ high.
  The subjects were mostly hosted in the ``Nest'' aviary, but transported to other aviaries (with their consent) to allow them to fly on slightly larger distances (6m), getting sun exposure, access distinct physical games and more generally to vary their stimuli.
The sessions of the development, training and testing phases were almost always realized in a training area next to the opening of the ``Nest'' aviary, and in a few occasions insider the larger ``North'' aviary.
\end{LONG}
At no point were the subjects food or water deprived: at any point they could fly to their housing space, where food and water was available.
\begin{LONG}
The sessions always occurred on one of three similar wood frames (see Figure~\ref{fig:stableSetup}), so that to offer a familiar setting even when the location of the training changed (e.g. in the ``North'' aviary).
Even though the digital devices had to be replaced at some point, those were always hold on the same wood structure (etched with the name of the subject to which it was assigned), so that to facilitate the recognition of which device was assigned to which subject.
The subjects were weighted on a weekly basis to detect any variation which could indicate a potential health issue, and brought to a licensed veterinarian twice a year.

\begin{figure}
\centering
\begin{minipage}[b]{.45\linewidth}
\includegraphics[width=\textwidth]{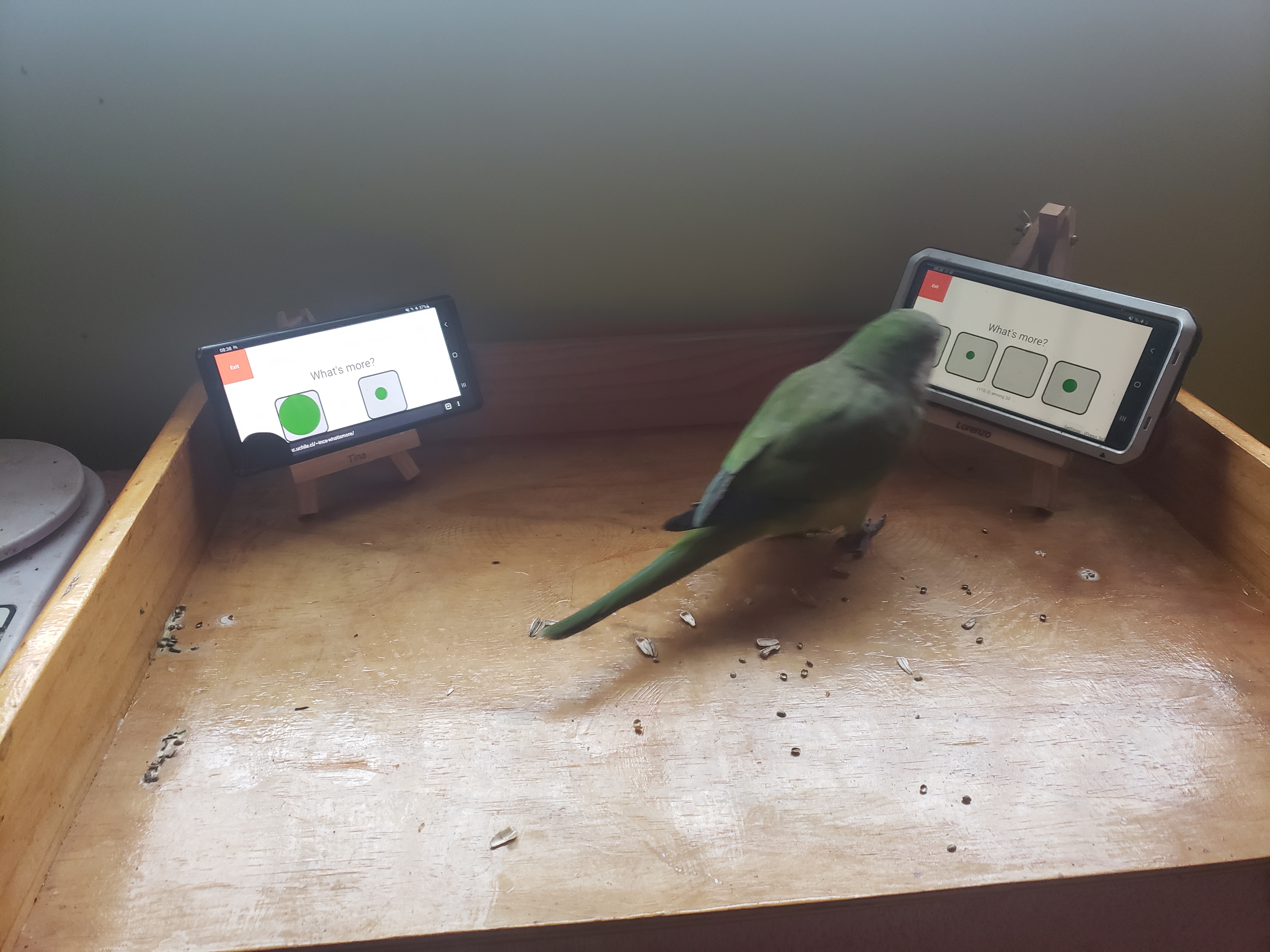}
  \Description{Picture of the experimental set-up in the learning phase, with two devices side by side and one Monk Parrot interacting with the right one.}
\caption{Each subject disposes of its own device, placed on wood supports of distinct sizes, each etched with the name of the subject to which it is assigned.\label{fig:plankSetup}}
\end{minipage}\hfill
\begin{minipage}[b]{.45\linewidth}
\includegraphics[width=\textwidth]{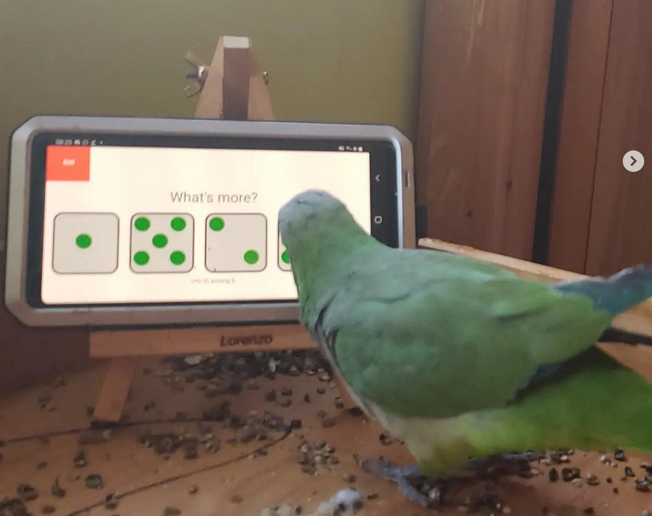}
\Description{Picture of a Monk Parakeet interacting with its assigned device, on a support at its name.}
  \caption{Each device is placed on a wood support (as opposed to carried by the experimenter), at a height comfortable to the subject, making the subject as autonomous as possible.\label{fig:stableSetup}}
\end{minipage}
\end{figure}
\end{LONG}

\subsubsection{Application Usability}\label{sec:appl-usab}
\begin{MAYB}
Summarize \citet{2019-ACI-AMethodForEvaluatingAnimalUsability-RugeMancini}'s rules.
\end{MAYB}
In order to minimize the potential frustration of the subjects when facing inadequate answers from the application, each version of the application was systematically tested by two human subjects, and any issue detected during such a phase corrected, before being presented to the nonhuman subjects.
During the phase of software development, when a feature of the application (whether due to an error or to an setting proved to be inadequate) was encountered to frustrate the subjects, the use of this application was replaced by another activity until the software was corrected, tested and separately validated by two human subjects.

\subsubsection{Sense of Agency}\label{sec:sense-agency}
\begin{HEADPARAGRAPH}
Both physical and virtual aspects of the protocol were designed so that to maintain a sense of agency in the subjects.
\end{HEADPARAGRAPH}
    \begin{MAYB}
READ and maybe mention the last (April 2022) article from 
\citet{2022-FAS-RelevanceImpartialityWelfareAndConsentPrinciplesOfAnAnimalCenteredResearchEthics-ManciniNannoni}
about consent
    \end{MAYB}
    The physical setting of the experimentation was designed so that to insure that the subject's participation was voluntary during all three phases of the process:
  the subjects were invited to come to the training area (but could, and sometime did, refuse);
  at any time the subjects could fly from the training area
  back to their aviary, 
  to a transportation pack with a large amount of seeds suspended above the training area, or
  to an alternate training area on the side, presenting an alternate choice of training exercises. 
%
Concerning the psychological aspects, 
  the main menu of the application was designed so that each subject can choose in which visualisation mode they wish to play (see Figures~\ref{fig:screenshotOfStartingMenu} and~\ref{fig:LorenzoChoosingModeInWhatIsMore}), and
  a large orange ``exit'' button is present on the playing screen allowing the subject to signal that they do not wish to play this game any more, prompting the experimenter to propose alternatives.

\begin{LONG}
  The page to adjust parameters controlling the difficulty (e.g. domain and number of values displayed, length of a game, etc.) of the games, more complex display and sound choices (e.g. colors and spaces being used in the display, words pronounced by the software in various situations, etc.), and the details about the application logs, is accessed via a special button requiring a longer press, making it harder to access to nonhuman subjects.
  \end{LONG}

\begin{MAYB}
\subsubsection{Nonhuman Privacy issues}\label{sec:nonh-priv-issu}
Summarize     \cite{2022-FVS-TheCaseForAnimalPrivacyInTheDesignOfTechnologicallySupportedEnvironments-PaciManciniNuseibeh}
and justify that the log data does not constitute information which could be inferred to be considered private by the nonhuman subjects.
\end{MAYB}

\subsubsection{Approval of the experimental protocol by CICUA}\label{sec:appr-exper-prot}
All interactions with animals were governed by a protocol reviewed and approved by the Institutional Animal Care and Use Committee (IACUC) \begin{NONANONYMOUS} (in Spanish ``COMITÉ INSTITUCIONAL de CUIDADO y USO de ANIMALES (CICUA)'') \end{NONANONYMOUS} of the researchers' institution, through a form of Experimentation Protocol of Management and Care of Animals\begin{NONANONYMOUS} (``Protocolo de Manejo y Cuidado de Animales'')\end{NONANONYMOUS}.

\subsection{Statistical Analysis Process}
\label{sec:statisticalAnalysis}

\begin{HEADPARAGRAPH}
The statistical analysis of the experimental results was designed as part of the experimental protocol, with the objectives to compute the accuracy of each subject for each display mode and each size of the set of values presented to the subject (Section \ref{sec:statisticalTools}), to compare it with the accuracy of selecting a value uniformly at random (Section~\ref{sec:binomialTests}) and to search for correlation between the answer's accuracy and some measure on the values presented (Section~\ref{sec:Pearson}).
\end{HEADPARAGRAPH}

\subsubsection{Statistical tools used}
\label{sec:statisticalTools}

The statistical analysis was performed in a python notebook, executed and shared via the collaborative website \url{https://colab.research.google.com}\begin{LONG}: this platform was chosen because it is easy to collaborate among peers as well as to run and replicate statistical analyses\end{LONG}.  Such python notebook was developed and tested on the logs generated during the (masked and unmasked) training sessions, to be used later without major modification on the logs generating during the masked experimental sessions of the testing phase.
\begin{LONG}
The computation made use of the following libraries:
\begin{itemize}
\item\texttt{pandas} is a library written as an extension of \texttt{numpy} to facilitate the manipulation and analysis of data.
\item\texttt{seaborn} and \texttt{matplotlib} are libraries for the visualisation of statistical data. \texttt{seaborn} was used for the creation of correlation graphs and \texttt{matplotlib} for heat maps.
\item\texttt{scipy} is a free and open source library for Python. It consists of mathematical tools and algorithms,  from this library we use \texttt{scipy.stats} for the chi-square and binomial tests.
\end{itemize}
The python notebook operates on the log files via the  \texttt{pandas} library.
These logs can be worked individually or concatenated to obtain a large overall analysis of the test subject.
\end{LONG}

\subsubsection{Binomial Tests}
\label{sec:binomialTests}

The average accuracy of each subject for each display mode and each size of the set of values presented to the subject is then the average of the \texttt{Correction} entry in the log (replacing \texttt{True} by $1$ and \texttt{False} by $0$) over all data points matching the criteria.
For each such accuracy, we performed a \texttt{binomial test} in order to decide if such accuracy was substantially better than that achieved by selecting a value uniformly at random.
\begin{LONG}
To calculate the binomial test we count the "success" among all the points of the dataset, and apply the \texttt{binom\_test} method in \texttt{scipy}.  $ p = binom\_test(k,n,prob,alternative='greater')$, where $k$ is the total number of successes, $n$ is the total number of attempts,
over tests selecting the maximal value out of two, $prob=0.5$;
over tests selecting the maximal value out of three,  $prob=0.33$; and
over tests selecting the maximal value out of four, $prob=0.25$.
The \textit{greater}  alternative is used since we are looking for an accuracy greater or equal to $50\%$ , $33\%$ and $25\%$ respectively. 
\end{LONG}
We performed such statistical analysis on the data of each particular session and on their union, on each particular visualization mode and on the type of visualisation mode (discrete or continuous) and on all visualisation modes (see Tables~\ref{tab:finerAnalysisTableMaxOutOfTwoSubject1}, \ref{tab:finerAnalysisTableMaxOutOfTwoSubject2}, \ref{tab:finerAnalysisTableMaxOutOfThree} and~\ref{tab:finerAnalysisTableMaxOutOfFour}).

\subsubsection{Pearson Correlation Analysis}
\label{sec:Pearson}

In order to compare our results with that of \citet{2008-AC-TheDiscriminationOfDiscreteAndContinuousAmountsInAfricanGreyParrots-AlAinGiretGrandKreutzerBovet}'s experiments, we performed a Pearson correlation analysis of the relation between the accuracy of the subjects' answers when asked to select the maximal out of two values on one hand, and the three variables they considered on the other hand:
\begin{itemize}
\item the \emph{sum} of the values for each test (e.g. from $1+2=3$ to $4+5=9$),
\item the \emph{difference} between the two extreme values presented within a trial (e.g. from $1$ to $5-1=4$) and
\item the \emph{ratio} of continuous quantities presented, by dividing the smallest presented value by the largest one (e.g. from $\frac{1}{5}=0.2$ to $\frac{4}{5}=0.8$).
  \end{itemize}

  \begin{BRIDGEPARAGRAPH}
We describe the results of the experiments and their statistical analysis in the next section.
\end{BRIDGEPARAGRAPH}

\section{Results}
\label{sec:results}

After relatively long phases of development and training (15 months) using various domains of values (from $\{0,1\}$ to $\{0,1,\ldots,9\}$), the experimental phase was quite short (one week), with all experiments performed using a masked setup and a domain of values restricted to the set $\{1,2,3,4,5\}$ in order to stay as close as possible to the settings of \citet{2008-AC-TheDiscriminationOfDiscreteAndContinuousAmountsInAfricanGreyParrots-AlAinGiretGrandKreutzerBovet}'s study.
\begin{HEADPARAGRAPH}
We summarize the number and content of the logs obtained (Section~\ref{sec:logResults}),
perform binomial tests on the experimental results when choosing the maximal value out of two for both subjects (Section~\ref{sec:maxOutOfTwo}),
perform binomial tests on the experimental results when choosing the maximal value out of three and four for the first subject (Section~\ref{sec:maxOutOfThreeAndFour}), and perform various correlation tests between the performance of the subjects and the \emph{sum}, \emph{difference} and \emph{ratio} of the values presented (Section~\ref{sec:relat-betw-perf}).
\end{HEADPARAGRAPH}
\begin{LONG}

\subsection{Log results}
\label{sec:logResults}

\end{LONG}
A testing session typically lasts some 5 to 10 games of 20 questions each, resulting into a log of 100 to 200 data points: see Figures~\ref{fig:shortLogExampleVerbatim} and \ref{fig:shortExampleOfLog} for a shortened example of log\begin{LONG}, and Figure~\ref{fig:pageLongLogExample} for a longer one\end{LONG}.
\begin{LONG}
\begin{figure}
\centering
\begin{adjustbox}{max width=\linewidth}
\begin{BVerbatim}
Test no, Test Name, Learner, Trainer, C_0, C_1, C_2, C_3, C_4, Value selected , Correction , Date, Answering Time (ms), Other Parameters
1, dice, Subject, Experimenter, 1,4,,,, 4,true, [2022-05-19 17:02(25.981)], 7946, background black, foreground green, bg opacity .2, Value Set [1,2,3,4,5]
2, dice, Subject, Experimenter, 1,5,,,, 5,true, [2022-05-19 17:02(30.82)], 3095, background black, foreground green, bg opacity .2, Value Set [1,2,3,4,5]
3, dice, Subject, Experimenter, 3,4,,,, 4,true, [2022-05-19 17:02(39.70)], 7981, background black, foreground green, bg opacity .2, Value Set [1,2,3,4,5]
4, dice, Subject, Experimenter, 1,4,,,, 4,true, [2022-05-19 17:02(46.295)], 6217, background black, foreground green, bg opacity .2, Value Set [1,2,3,4,5]
5, dice, Subject, Experimenter, 2,5,,,, 5,true, [2022-05-19 17:02(51.633)], 4331, background black, foreground green, bg opacity .2, Value Set [1,2,3,4,5]
6, dice, Subject, Experimenter, 3,1,,,, 1,false, [2022-05-19 17:03(00.79)], 7440, background black, foreground green, bg opacity .2, Value Set [1,2,3,4,5]
7, dice, Subject, Experimenter, 4,1,,,, 1,false, [2022-05-19 17:03(02.938)], 1852, background black, foreground green, bg opacity .2, Value Set [1,2,3,4,5]
8, dice, Subject, Experimenter, 1,2,,,, 2,true, [2022-05-19 17:03(06.86)], 2141, background black, foreground green, bg opacity .2, Value Set [1,2,3,4,5]
9, dice, Subject, Experimenter, 3,2,,,, 2,false, [2022-05-19 17:03(13.478)], 6383, background black, foreground green, bg opacity .2, Value Set [1,2,3,4,5]
10, dice, Subject, Experimenter, 4,5,,,, 5,true, [2022-05-19 17:03(16.578)], 2094, background black, foreground green, bg opacity .2, Value Set [1,2,3,4,5]
11, dice, Subject, Experimenter, 4,5,,,, 5,true, [2022-05-19 17:03(20.412)], 2826, background black, foreground green, bg opacity .2, Value Set [1,2,3,4,5]
12, dice, Subject, Experimenter, 1,4,,,, 4,true, [2022-05-19 17:03(28.740)], 7321, background black, foreground green, bg opacity .2, Value Set [1,2,3,4,5]
13, dice, Subject, Experimenter, 1,2,,,, 2,true, [2022-05-19 17:03(40.376)], 10629, background black, foreground green, bg opacity .2, Value Set [1,2,3,4,5]
14, dice, Subject, Experimenter, 1,2,,,, 2,true, [2022-05-19 17:03(53.7)], 11624, background black, foreground green, bg opacity .2, Value Set [1,2,3,4,5]
15, dice, Subject, Experimenter, 4,2,,,, 2,false, [2022-05-19 17:03(57.33)], 3018, background black, foreground green, bg opacity .2, Value Set [1,2,3,4,5]
16, dice, Subject, Experimenter, 5,1,,,, 1,false, [2022-05-19 17:04(00.23)], 1984, background black, foreground green, bg opacity .2, Value Set [1,2,3,4,5]
17, dice, Subject, Experimenter, 2,3,,,, 3,true, [2022-05-19 17:04(03.156)], 2127, background black, foreground green, bg opacity .2, Value Set [1,2,3,4,5]
18, dice, Subject, Experimenter, 4,1,,,, 1,false, [2022-05-19 17:04(07.608)], 3443, background black, foreground green, bg opacity .2, Value Set [1,2,3,4,5]
19, dice, Subject, Experimenter, 4,3,,,, 3,false, [2022-05-19 17:04(11.969)], 3354, background black, foreground green, bg opacity .2, Value Set [1,2,3,4,5]
20, dice, Subject, Experimenter, 2,3,,,, 3,true, [2022-05-19 17:04(16.363)], 3386, background black, foreground green, bg opacity .2, Value Set [1,2,3,4,5]
21, rect, Subject, Experimenter, 4,5,,,, 5,true, [2022-05-19 17:04(31.694)], 7191, background black, foreground green, bg opacity .2, Value Set [1,2,3,4,5]
22, rect, Subject, Experimenter, 3,4,,,, 4,true, [2022-05-19 17:04(36.330)], 3629, background black, foreground green, bg opacity .2, Value Set [1,2,3,4,5]
23, rect, Subject, Experimenter, 5,1,,,, 1,false, [2022-05-19 17:04(43.148)], 5810, background black, foreground green, bg opacity .2, Value Set [1,2,3,4,5]
24, rect, Subject, Experimenter, 4,2,,,, 2,false, [2022-05-19 17:04(44.926)], 772, background black, foreground green, bg opacity .2, Value Set [1,2,3,4,5]
25, rect, Subject, Experimenter, 1,5,,,, 5,true, [2022-05-19 17:04(46.731)], 798, background black, foreground green, bg opacity .2, Value Set [1,2,3,4,5]
26, rect, Subject, Experimenter, 1,5,,,, 5,true, [2022-05-19 17:04(51.117)], 3380, background black, foreground green, bg opacity .2, Value Set [1,2,3,4,5]
27, rect, Subject, Experimenter, 1,5,,,, 5,true, [2022-05-19 17:04(55.855)], 3730, background black, foreground green, bg opacity .2, Value Set [1,2,3,4,5]
28, rect, Subject, Experimenter, 1,4,,,, 4,true, [2022-05-19 17:05(00.581)], 3718, background black, foreground green, bg opacity .2, Value Set [1,2,3,4,5]
29, rect, Subject, Experimenter, 3,2,,,, 3,true, [2022-05-19 17:05(05.74)], 3487, background black, foreground green, bg opacity .2, Value Set [1,2,3,4,5]
30, rect, Subject, Experimenter, 2,4,,,, 4,true, [2022-05-19 17:05(08.885)], 2803, background black, foreground green, bg opacity .2, Value Set [1,2,3,4,5]
31, rect, Subject, Experimenter, 3,5,,,, 3,false, [2022-05-19 17:05(16.709)], 6818, background black, foreground green, bg opacity .2, Value Set [1,2,3,4,5]
32, rect, Subject, Experimenter, 4,1,,,, 4,true, [2022-05-19 17:05(18.396)], 678, background black, foreground green, bg opacity .2, Value Set [1,2,3,4,5]
33, rect, Subject, Experimenter, 4,2,,,, 2,false, [2022-05-19 17:05(22.709)], 3306, background black, foreground green, bg opacity .2, Value Set [1,2,3,4,5]
34, rect, Subject, Experimenter, 5,3,,,, 3,false, [2022-05-19 17:05(25.95)], 1380, background black, foreground green, bg opacity .2, Value Set [1,2,3,4,5]
35, rect, Subject, Experimenter, 5,2,,,, 5,true, [2022-05-19 17:05(29.448)], 3347, background black, foreground green, bg opacity .2, Value Set [1,2,3,4,5]
(...)
145, disc, Subject, Experimenter, 4,2,1,,, 2,false, [2022-05-19 17:32(07.287)], 802, background black, foreground green, bg opacity .2, Value Set [1,2,3,4,5]
146, disc, Subject, Experimenter, 5,4,3,,, 4,false, [2022-05-19 17:32(26.796)], 18496, background black, foreground green, bg opacity .2, Value Set [1,2,3,4,5]
147, disc, Subject, Experimenter, 4,5,1,,, 5,true, [2022-05-19 17:32(32.142)], 4334, background black, foreground green, bg opacity .2, Value Set [1,2,3,4,5]
148, disc, Subject, Experimenter, 3,5,4,,, 5,true, [2022-05-19 17:32(43.167)], 10015, background black, foreground green, bg opacity .2, Value Set [1,2,3,4,5]
149, disc, Subject, Experimenter, 5,4,1,,, 5,true, [2022-05-19 17:32(51.628)], 7448, background black, foreground green, bg opacity .2, Value Set [1,2,3,4,5]
150, disc, Subject, Experimenter, 4,2,5,,, 5,true, [2022-05-19 17:32(55.922)], 3283, background black, foreground green, bg opacity .2, Value Set [1,2,3,4,5]
151, disc, Subject, Experimenter, 5,2,3,,, 5,true, [2022-05-19 17:33(00.783)], 3850, background black, foreground green, bg opacity .2, Value Set [1,2,3,4,5]
152, disc, Subject, Experimenter, 1,4,2,,, 4,true, [2022-05-19 17:33(05.553)], 3757, background black, foreground green, bg opacity .2, Value Set [1,2,3,4,5]
153, disc, Subject, Experimenter, 5,1,4,,, 4,false, [2022-05-19 17:33(08.485)], 1921, background black, foreground green, bg opacity .2, Value Set [1,2,3,4,5]
154, disc, Subject, Experimenter, 5,1,3,,, 5,true, [2022-05-19 17:33(10.272)], 778, background black, foreground green, bg opacity .2, Value Set [1,2,3,4,5]
155, disc, Subject, Experimenter, 1,5,3,,, 5,true, [2022-05-19 17:33(15.484)], 4204, background black, foreground green, bg opacity .2, Value Set [1,2,3,4,5]
156, disc, Subject, Experimenter, 5,4,3,,, 4,false, [2022-05-19 17:33(19.152)], 2658, background black, foreground green, bg opacity .2, Value Set [1,2,3,4,5]
157, disc, Subject, Experimenter, 4,2,3,,, 3,false, [2022-05-19 17:33(22.15)], 1853, background black, foreground green, bg opacity .2, Value Set [1,2,3,4,5]
158, disc, Subject, Experimenter, 1,5,3,,, 5,true, [2022-05-19 17:33(24.137)], 1111, background black, foreground green, bg opacity .2, Value Set [1,2,3,4,5]
159, disc, Subject, Experimenter, 2,3,5,,, 5,true, [2022-05-19 17:33(27.386)], 2240, background black, foreground green, bg opacity .2, Value Set [1,2,3,4,5]
160, disc, Subject, Experimenter, 3,1,4,,, 3,false, [2022-05-19 17:33(30.852)], 2456, background black, foreground green, bg opacity .2, Value Set [1,2,3,4,5]
161, heap, Subject, Experimenter, 5,1,3,,, 5,true, [2022-05-19 17:33(40.558)], 3063, background black, foreground green, bg opacity .2, Value Set [1,2,3,4,5]
162, heap, Subject, Experimenter, 4,1,2,,, 4,true, [2022-05-19 17:33(46.447)], 4879, background black, foreground green, bg opacity .2, Value Set [1,2,3,4,5]
163, heap, Subject, Experimenter, 3,4,2,,, 4,true, [2022-05-19 17:33(51.684)], 4225, background black, foreground green, bg opacity .2, Value Set [1,2,3,4,5]
164, heap, Subject, Experimenter, 2,1,4,,, 4,true, [2022-05-19 17:33(54.658)], 1963, background black, foreground green, bg opacity .2, Value Set [1,2,3,4,5]
165, heap, Subject, Experimenter, 2,3,4,,, 4,true, [2022-05-19 17:33(59.410)], 3741, background black, foreground green, bg opacity .2, Value Set [1,2,3,4,5]
166, heap, Subject, Experimenter, 2,1,3,,, 2,false, [2022-05-19 17:34(03.670)], 3247, background black, foreground green, bg opacity .2, Value Set [1,2,3,4,5]
167, heap, Subject, Experimenter, 5,4,3,,, 3,false, [2022-05-19 17:34(05.808)], 1128, background black, foreground green, bg opacity .2, Value Set [1,2,3,4,5]
168, heap, Subject, Experimenter, 2,3,1,,, 3,true, [2022-05-19 17:34(07.846)], 1029, background black, foreground green, bg opacity .2, Value Set [1,2,3,4,5]
169, heap, Subject, Experimenter, 2,3,5,,, 5,true, [2022-05-19 17:34(19.932)], 11078, background black, foreground green, bg opacity .2, Value Set [1,2,3,4,5]
170, heap, Subject, Experimenter, 2,3,4,,, 4,true, [2022-05-19 17:34(25.404)], 4462, background black, foreground green, bg opacity .2, Value Set [1,2,3,4,5]
171, heap, Subject, Experimenter, 1,5,2,,, 5,true, [2022-05-19 17:34(31.359)], 4945, background black, foreground green, bg opacity .2, Value Set [1,2,3,4,5]
172, heap, Subject, Experimenter, 1,4,5,,, 5,true, [2022-05-19 17:34(35.552)], 3182, background black, foreground green, bg opacity .2, Value Set [1,2,3,4,5]
173, heap, Subject, Experimenter, 2,4,3,,, 4,true, [2022-05-19 17:34(39.311)], 2747, background black, foreground green, bg opacity .2, Value Set [1,2,3,4,5]
174, heap, Subject, Experimenter, 2,4,3,,, 4,true, [2022-05-19 17:34(44.188)], 3867, background black, foreground green, bg opacity .2, Value Set [1,2,3,4,5]
175, heap, Subject, Experimenter, 5,3,1,,, 5,true, [2022-05-19 17:34(48.508)], 3308, background black, foreground green, bg opacity .2, Value Set [1,2,3,4,5]
176, heap, Subject, Experimenter, 3,4,2,,, 4,true, [2022-05-19 17:34(53.468)], 3950, background black, foreground green, bg opacity .2, Value Set [1,2,3,4,5]
177, heap, Subject, Experimenter, 3,2,5,,, 5,true, [2022-05-19 17:34(56.511)], 2032, background black, foreground green, bg opacity .2, Value Set [1,2,3,4,5]
178, heap, Subject, Experimenter, 5,2,4,,, 4,false, [2022-05-19 17:35(02.323)], 4802, background black, foreground green, bg opacity .2, Value Set [1,2,3,4,5]
179, heap, Subject, Experimenter, 5,2,4,,, 4,false, [2022-05-19 17:35(04.70)], 737, background black, foreground green, bg opacity .2, Value Set [1,2,3,4,5]
180, heap, Subject, Experimenter, 3,2,1,,, 2,false, [2022-05-19 17:35(06.6)], 926, background black, foreground green, bg opacity .2, Value Set [1,2,3,4,5]
\end{BVerbatim}
\end{adjustbox}
\caption{A page long example of log generated by the application, obtained by taking the 35 first and last entries of the log of the masked session on the afternoon of the 19th of May 2022. One can observe that the session started with two games selecting the largest of two values, displayed as dice and as rectangles, and ended with two games selecting the largest of three values, displayed as discs and heaps. \label{fig:pageLongLogExample}}
\end{figure}
\end{LONG}
The testing phase occurred between the 19th of May 2022 and the 26th of May 2022.
These experiments used four different display modes (``Dice'', ``Heap'', ``Disc'' and ``Rectangle''), requesting the subject to select the maximal value out of a set of 2, 3 or 4 values, randomly chosen among a set of five values $\{1,2,3,4,5\}$, in order to produce a setup relatively similar to that of \citet{2008-AC-TheDiscriminationOfDiscreteAndContinuousAmountsInAfricanGreyParrots-AlAinGiretGrandKreutzerBovet}, with the vast majority of experiments selecting the maximal out of two values, and only a few out of three or four values. Each log corresponds to a separate training session and device, containing between 80 and 400 entries (each entry being a separate question and answer).
In total, 14 logs were collected for the first subject, and 5 logs were collected for the second subject: the first subject was requested to select the maximal value out of 2,3 or 4 values, while the second subject was requested to select the maximal value only out of 2 values. 
\begin{LONG}
Concerning the selection of the maximal out of 2 values (the setting the most similar to that of \citet{2008-AC-TheDiscriminationOfDiscreteAndContinuousAmountsInAfricanGreyParrots-AlAinGiretGrandKreutzerBovet}), the first subject answered 449 dice tests, 400 heap tests, 262 rectangle tests and 103 disc tests, making a total of 1214 tests, while the second subject answered 190 dice tests 26 rectangle tests and 193 disc tests making a total of 409 tests.
Concerning the selection of the maximal out of 3 values, the first subject answered 249 dice tests, 120 heap tests,  120 rectangle tests, and 99 disc tests, making a total of 588 tests.
Concerning the selection of the maximal out of 4 values, the first subject answered
154 dice tests, 51 heap tests, and 13 disc tests, making a total of 218 tests.

\end{LONG}
See Table~\ref{tab:numberOfDataPoints} for a summary of the number of data points collected separated by display modes (``Dice'', ``Heap'', ``Disc'' and ``Rectangle''), accumulated by the type of display mode (``Discrete'' or ``Continuous'') and accumulated over all display modes (``Total'').
\begin{LONG}
Even though the care to respect the agency of the subjects introduced great imbalances between the number of data points collected for each display mode and set size, 2429 data points were gathered in only one week, through voluntary participation in the subject: this is much higher than what could be achieved in the same amount of time via traditional analogical protocols such that of \citet{2008-AC-TheDiscriminationOfDiscreteAndContinuousAmountsInAfricanGreyParrots-AlAinGiretGrandKreutzerBovet}.
\end{LONG}
\begin{table}
\centering
\begin{tabular}{c|c*{2}{||r|r||r}||r}
Subject & Set Size & Dice & Heap & Discrete & Disc & Rectangle & Continuous & Total \\
  \hline
  1     & 2        & 449  & 400  & 849      & 103  & 262       & 448        & 1214  \\
  1     & 3        & 249  & 120  & 0        & 126  & 120       & 0          & 588   \\
  1     & 4        & 154  & 51   & 205      & 13   & 0         & 13         & 218   \\
  2     & 2        & 190  & 0    & 190      & 193  & 26        & 219        & 409   \\
  \hline
  1     & total    & 852  & 571  & 1054     & 242  & 382       & 461        & 2020  \\
  2     & total    & 190  & 0    & 190      & 193  & 26        & 219        & 409   \\
  \hline
  total & total    & 1042 & 644  &  1244   & 435  & 468       & 680        & 2429  \\
\end{tabular}
\caption{Number of data points collected separated by display modes (``Dice'', ``Heap'', ``Disc'' and ``Rectangle''), accumulated by the type of display mode (``Discrete'' or ``Continuous'') and accumulated over all display modes (``Total'').
  The imbalance between the frequencies of the display modes and between the amounts of test results for each subjects is explained by the care to support the agency of the subjects: they could interrupt the session at any time, and had the option to choose the display mode at any time (even though they seldom did).  }
\label{tab:numberOfDataPoints}
\end{table}
\begin{BRIDGEPARAGRAPH}
We analyze those results statistically in the following sections.
\end{BRIDGEPARAGRAPH}

\subsection{Selecting the maximal value out of two}
\label{sec:maxOutOfTwo}
\begin{HEADPARAGRAPH}
Both subjects played the game in the four display modes, the first subject showing much more interest in participating than the second one, but none of them marking a particular preference for any display mode.  The first subject showed an average accuracy of $81.79\%$ (Section~\ref{sec:subject1}), the second subject an average accuracy of $74\%$ (Section~\ref{sec:subject2}). Both performed better when the values were very different and worse when the values were close (Section~\ref{sec:relat-betw-perf}), exactly as the three African Grey parrots in \cite{2008-AC-TheDiscriminationOfDiscreteAndContinuousAmountsInAfricanGreyParrots-AlAinGiretGrandKreutzerBovet}'s study.
\end{HEADPARAGRAPH}

\subsubsection{First Subject}
\label{sec:subject1}

The results show a clear ability from the first subject to discriminate the maximal value out of two quantities.
Over all experimentations requesting to select the maximal value out of two, the first subject responded correctly $993$ times out of a total of $1214$ trials, corresponding to an average accuracy of $81.79\%$.
A simple binomial tests indicates that the probability to achieve such an accuracy by answering uniformly at random $1214$ such binary questions is $p=1.95 \cdot 10^{-117}$ (see Table~\ref{tab:finerAnalysisTableMaxOutOfTwoSubject1} for a more detailed description of the results, by session and by display mode).
\begin{LONG}
It is very likely that the subject did \emph{not} answer uniformly at random.
Lacking a bias in the experimental protocol, this seems to indicate a clear ability to discriminate between the two values being presented.

\end{LONG}

\begin{LONG}
Analyzing separately the results for each display mode or type of display mode, corroborates the result and points out some interesting facts:
First, one does not observe any relevant improvement over time, which is explained by the relatively long period of training before the relatively short period of testing.
Second, over all the subject performed with a slightly better accuracy for continuous display modes ($88\%$ vs $79\%$) and, surprisingly (because one would expect the reverse for human, and similar accuracy for nonhumans), a much better accuracy for the ``Heap'' display mode over the ``Dice'' display mode ($85\%$ vs $74\%$).
\end{LONG}

\begin{table}
\centering
 \begin{tabular}{c*{2}{||c|c||c}||c}
Session        & Dice            & Heap             & Discrete        & Disc             & Rectangle       & Continuous      & Total            \\
\hline
        19,17h & $65 (1e^{-1})$  & $60 (2e^{-1})$   & $62 (7e^{-2})$  & $90 (2e^{-4})$   & $75 (2e^{-2})$  & $82 (2e^{-5})$  & $72 (3e^{-5})$   \\ 
        21,17h & $80 (5e^{-17})$ & $93 (1e^{-15})$  & $84 (1e^{-29})$ & $91 (3e^{-5})$   & $95 (3e^{-14})$ & $94 (3e^{-18})$ & $86 (6e^{-45})$  \\ 
        23,08h & $80 (1e^{-5})$  & $84 (2e^{-3})$   & $81 (8e^{-8})$  & (no data)        & $90 (8e^{-15})$ & $90 (8e^{-15})$ & $86 (1e^{-20})$  \\ 
        23,15h & $70 (5e^{-3})$  & $86 (1e^{-20})$  & $82 (8e^{-21})$ & $88 (1e^{-8})$   & (no data)       & $88 (1e^{-8})$  & $83 (1e^{-27})$  \\ 
        24,10h & $66 (1e^{-4})$  & (no data)        & $66 (1e^{-4})$  & (no data)        & (no data)       & (no data)       & $66 (1e^{-4})$   \\ 
        24,17h & (no data)       & $83 (3e^{-16})$  & $83 (3e^{-16})$ & (no data)        & (no data)       & (no data)       & $83 (3e^{-16})$  \\ 
        25,08h & (no data)       & (no data)        & (no data)       & $60 (3e^{-1})$   & $86 (4e^{-14})$ & $84 (1e^{-13})$ & $84 (1e^{-13})$  \\ 
        25,13h & $71 (3e^{-2})$  & (no data)        & $71 (3e^{-2})$  & (no data)        & (no data)       & (no data)       & $71 (3e^{-2})$   \\ 
  \hline
       Total   & $74 (3e^{-25})$ & $85 (78e^{-49})$ & $79 (6e^{-69})$ & $86 (81e^{-15})$ & $89 (3e^{-40})$ & $88 (2e^{-53})$ & $82 (1e^{-117})$ \\ \hline
    \end{tabular}
    \caption{Finer analysis of the first subject's performance on selecting the maximal value out of two, separated by display modes (``Dice'', ``Heap'', ``Disc'' and ``Rectangle''), accumulated by the type of display mode (``Discrete'' or ``Continuous'') and accumulated over all display modes (``Total'').
      The sessions occurred during the month of May 2022 and are identified by the date $d$ and hour $h$ (e.g. the session which occurred at 17:02 on the 19th of May 2022 is identified by the tag ``19,17h''). Each entry is in the format $a(p)$ where $a$ is the accuracy reported, and $p$ is the probability of achieving such accuracy or better by selecting answers uniformly at random.
      Note how the accuracy percentages are mostly above $80\%$, and that the probability of such accuracy or a better one to be attained by selecting answers uniformly at random is smaller than $0.001$ in almost all the cases.  }
\label{tab:finerAnalysisTableMaxOutOfTwoSubject1}
\end{table}

\subsubsection{Second Subject}
\label{sec:subject2}

The second subject was more reluctant to play, but showed a similar ability.
Overall experimentations requesting to select the maximal value out of two during the testing phase, the second subject responded correctly $303$ times out of a total of $409$ trials, corresponding to an average accuracy of $74\%$.  A simple binomial tests indicates that the probability of answering correctly $383$ or more such binary questions out of $409$ by answering uniformly at random is $p=2.24 \cdot 10^{-23}$\begin{LONG}: here again, it is very likely that the subject did \emph{not} answer uniformly at random. Lacking a bias in the experimental protocol, this seems to indicate a clear ability to discriminate between the two values being presented\end{LONG}.

\begin{table}
\centering
\begin{tabular}{c*{2}{||c|c|c}||c}
Session & Dice            & Heap      & Discrete        & Disc            & Rect           & Continuous      & Total           \\ \hline
21,10h  & $84 (5e^{-15})$ & (no data) & $84 (5e^{-15})$ & (no data)       & $73 (1e^{-2})$ & $73 (1e^{-2})$  & $82 (6e^{-16})$ \\ 
23,15h  & $64 (5e^{-2})$  & (no data) & $64 (5e^{-2})$  & (no data)       & (no data)      & (no data)       & $64 (5e^{-2})$  \\ 
24,08h  & (no data)       & (no data) & (no data)       & $79 (5e^{-13})$ & (no data)      & $79 (5e^{-13})$ & $79 (5e^{-13})$ \\ 
24,17h  & $51 (5e^{-1})$  & (no data) & $51 (5e^{-1})$  & $57 (2e^{-1})$  & (no data)      & $57 (2e^{-1})$  & $55 (2e^{-1})$  \\ 
\hline
Total   & $74 (3e^{-12})$ & (no data) & $74 (3e^{-12})$ & $73 (2e^{-11})$ & $73 (1e^{-2})$ & $73 (1e^{-12})$ & $74 (2e^{-23})$ \\
\end{tabular}
\caption{Finer analysis of the second subject's performance on selecting the maximal value out of two, separated by display mode and combined.
  Note how the accuracy percentages are in between 51\% (not much better than random, on the last session) and 84\% with an average of 74\% (both much better than random), and that the probability of such accuracy to be attained by selecting answers uniformly at random is $p < 0.001$ in almost all the cases.}
\label{tab:finerAnalysisTableMaxOutOfTwoSubject2}
\end{table}

\subsubsection{Relation between accuracy and variables}
\label{sec:relat-betw-perf}

When selecting the maximal value out of two, both subjects showed a lower accuracy when the two values were close (difference or ratio close to $1$): see Table~\ref{tab:percentOfCorrecResponsesSubject1} for the percentages of correct answers for each subject and each of the $10$ sets of values presented (ignoring the order).
Such results corroborate those of the three African Grey parrots in \citet{2008-AC-TheDiscriminationOfDiscreteAndContinuousAmountsInAfricanGreyParrots-AlAinGiretGrandKreutzerBovet}'s study.

\begin{figure}
\begin{minipage}[b]{.5\linewidth}
\includegraphics[width=\textwidth]{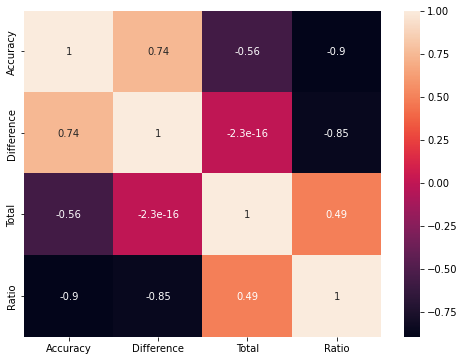}
\Description{Heat map correlation plot between the variables described in Table~\ref{tab:percentOfCorrecResponsesSubject1} for the first subject.}
\caption{Heat map correlation plot between the variables described in Table~\ref{tab:percentOfCorrecResponsesSubject1} for the first subject.
  Notice the strong negative correlation ($-0.9$) between \emph{Accuracy} and \emph{Ratio} one one hand, and the strong positive correlation ($0.74$) between \emph{Accuracy} and \emph{Difference} on the other hand.
 }
\label{fig:heatMapSubject1}
\end{minipage}\hfill\begin{minipage}[b]{.45\linewidth}
\includegraphics[width=\textwidth]{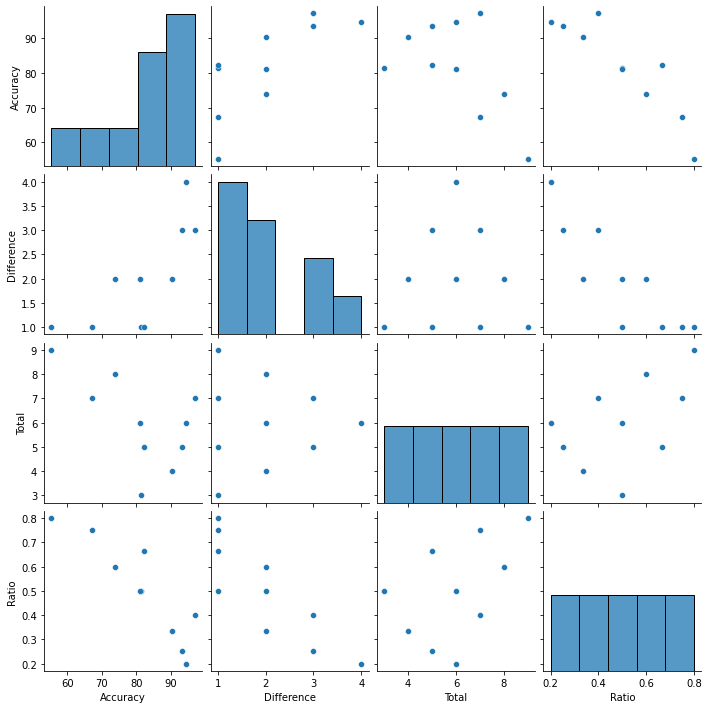}
\Description{Scatterplot between the variables described in Table~\ref{tab:percentOfCorrecResponsesSubject1} for the first subject.}
\caption{Scatter-plot of the variables described in Table~\ref{tab:percentOfCorrecResponsesSubject1} for the first subject.
  The diagonal plots show the distribution of the values of each variable.
  Note the uniform distribution of the \emph{Total} and \emph{Ratio}.
}
\label{fig:scatterPlotSubject1}
\end{minipage}
\end{figure}

Pearson's correlation tests for the first subject (see Figure~\ref{fig:heatMapSubject1} for the corresponding heat map and Figure~\ref{fig:scatterPlotSubject1} for the corresponding scatter plots) suggest an inverse correlation between the accuracy of the subject's selection and the ratio between the two values: for example, for a combination with small ratio $\frac{1}{5}=0.2$, the subject is more likely to correctly select the maximal value.
\begin{table}
    \centering
    \begin{tabular}{c|c|c|l|c|c}
Value Set & Total & Difference & Ratio & \multicolumn{2}{c}{Accuracy} \\
$(x,y)$   & $x+y$ & $y-x$      & $y/x$ & 1st Subject & 2nd Suject     \\ \hline
\{1,2\}   & 3     & 1          & 0.5   & 81\%        & 69\%           \\ 
\{1,3\}   & 4     & 2          & 0.33  & 90\%        & 70\%           \\ 
\{1,4\}   & 5     & 3          & 0.25  & 93\%        & 78\%           \\ 
\{1,5\}   & 6     & 4          & 0.2   & 94\%        & 94\%           \\ 
\{2,3\}   & 5     & 1          & 0.66  & 82\%        & 57\%           \\ 
\{2,4\}   & 6     & 2          & 0.5   & 81\%        & 68\%           \\ 
\{2,5\}   & 7     & 3          & 0.4   & 96\%        & 76\%           \\ 
\{3,4\}   & 7     & 1          & 0.75  & 67\%        & 45\%           \\ 
\{3,5\}   & 8     & 2          & 0.6   & 73\%        & 70\%           \\ 
\{4,5\}   & 9     & 1          & 0.8   & 55\%        & 71\%           \\ 
    \end{tabular}
    \caption{Both subject's \textit{Accuracy} for each pairs of values from the domain $\{1,2,3,4,5\}$.  %
      \textit{Total} is the total value of the representation shown to the test subject, \textit{Difference} is the difference between the two values presented, \textit{Ratio} is equal to the smallest quantity divided by the largest quantity.
      For the first subject, note how the lowest \emph{Accuracy} ($55\%$) corresponds to the highest \emph{ratio} ($0.8$), while for the second subject the lowest \emph{Accuracy} ($45\%$) corresponds to the second highest \emph{ratio} ($0.75$), suggesting a trend confirmed by the Pearson's correlation tests.}
    \label{tab:percentOfCorrecResponsesSubject1}
    \label{tab:percentOfCorrecResponsesSubject2}
\end{table}
There is a strong negative correlation ratio of $r=-0.9$ between the accuracy and the ratio, and a positive correlation ratio of $r=0.74$ between the accuracy and the difference (see the heat map in Figure~\ref{fig:heatMapSubject1}).
The scatter plots (in Figure~\ref{fig:scatterPlotSubject1}) show a decreasing relationship between the accuracy and the ratio, and an increasing relationship between the accuracy and the difference.  

\begin{figure}
\begin{minipage}[b]{.5\linewidth}
\includegraphics[width=\textwidth]{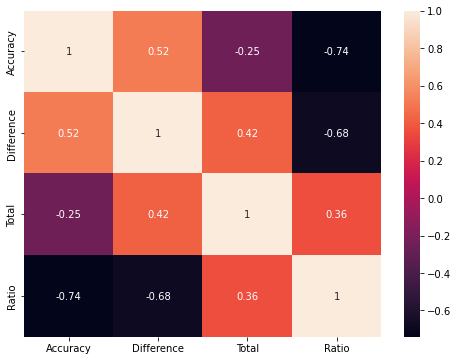}
\Description{Heat map correlation plot between the variables described in Table~\ref{tab:percentOfCorrecResponsesSubject2} for the second subject.}
\caption{Heat map correlation plot between the variables described in Table~\ref{tab:percentOfCorrecResponsesSubject2} for the second subject.
  Notice the negative correlation ($-0.74$) between \emph{Accuracy} and \emph{Ratio}.
}
\label{fig:heatMapSubject2}
\end{minipage}\hfill\begin{minipage}[b]{.45\linewidth}
\includegraphics[width=\textwidth]{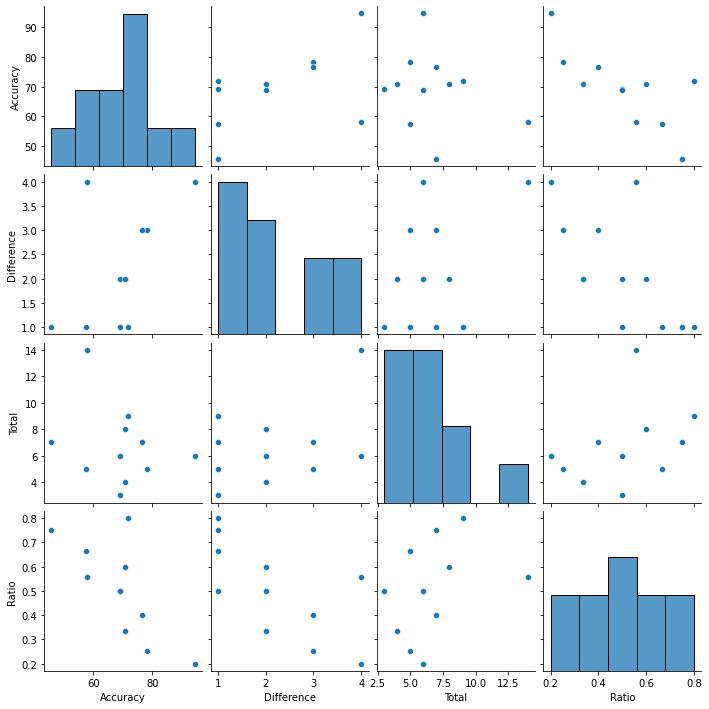}
\Description{Scatter plots correlation plot between the variables described in Table~\ref{tab:percentOfCorrecResponsesSubject2} for the second subject.}
\caption{Scatter plots for the variables described in Table~\ref{tab:percentOfCorrecResponsesSubject2} for the second subject.
  }
\label{fig:scatterPlotSubject2}
\end{minipage}
\end{figure}

There is a similar correlation between accuracy and ratio in the results of the second subject (see the heat-map in Figure~\ref{fig:heatMapSubject2} and the scatter plots in Figure~\ref{fig:scatterPlotSubject2}).
There is a strong negative correlation ratio of $r=-0.74$  between the ratio and the accuracy. The correlation ratio of $r=0.52$  between the difference and the accuracy is much weaker. 

\subsection{Selecting the maximal value out of three and four values}
\label{sec:maxOutOfThreeAndFour}

Only the first subject was tested on selecting the maximal value out of three and four values\begin{LONG}: the second subject  chose to stay in the ``Nest'' aviary or to play the digital piano for the remaining sessions\end{LONG}.
The subject a lower accuracy when asked to select the maximal value out of 3 or 4, than out of 2: on average the achieved an accuracy of $70\%$ for selecting the maximal out of three and $62\%$ for the maximal out of four, but still much better that what would be expected ($33\%$ and $25\%$ respectively) if the subject chose uniformly randomly among the values proposed (see Tables~\ref{tab:finerAnalysisTableMaxOutOfThree} and~\ref{tab:finerAnalysisTableMaxOutOfFour} for the detailed performances separated by display mode and sessions).
\begin{table}
\centering
\begin{tabular}{c*{2}{||c|c|c}||c}
  Session      & Dice            & Heap            & Discrete        & Disc            & Rect            & Continuous      & Total           \\ \hline
        19,17h & $55 (4e^{-3})$  & $75 (2e^{-4})$  & $61 (7e^{-6})$  & $6 (1e^{-2})$   & $85 (3e^{-6})$  & $72 (5e^{-7})$  & $66 (3e^{-11})$ \\ 
        22,09h & $51 (4e^{-3})$  & $65 (4e^{-4})$  & $56 (1e^{-5})$  & $78 (1e^{-10})$ & (no data)       & $78 (1e^{-10})$ & $64 (2e^{-13})$ \\ 
        22,11h & $57 (1e^{-4})$  & $64 (9e^{-6})$  & $60 (6e^{-9})$  & $90 (1e^{-16})$ & $89 (6e^{-31})$ & $89 (3e^{-46})$ & $77 (2e^{-47})$ \\ 
        25,16h & $67 (6e^{-12})$ & (no data)       & $67 (6e^{-12})$ & (no data)       & (no data)       & (no data)       & $67 (6e^{-12})$ \\ 
  \hline
        Total  & $59 (1e^{-17})$ & $66 (1e^{-11})$ & $61 (2e^{-27})$ & $80 (1e^{-25})$ & $88 (7e^{-36})$ & $84 (2e^{-59})$ & $70 (3e^{-77})$ \\ \
\end{tabular}
\caption{Finer analysis of the first subject's performance on selecting the maximal value out of three, separated by display mode and combined.
  Note how the average accuracy of random position selection in this case is 33\%, so an accuracy between 51\% and 84\% is a reasonable measure, as well as the probability  to achieve such accuracy or above when choosing one of the value between 3 options uniformly at random }
\label{tab:finerAnalysisTableMaxOutOfThree}
\end{table}
Two simple binomial tests give a more formal measure of how much better the subject performed compared to someone choosing uniformly at random: the probabilities of obtaining an accuracy equivalent or superior by randomly choosingthe same number of answers is $p = 3.479 \cdot 10^{-77}$ with probability $0.33$ of success (for selecting the maximal out of 3 values $588$ times) and $p = 2.549 \cdot 10^{-31}$ with probability $0.25$ of success (for selecting the maximal out of 4 values $136$ times): with very high probabilty, the subject showed their ability to discriminate between three and four values.

\begin{table}
\centering
\begin{tabular}{c*{2}{||c|c|c}||c}
  Session      & Dice            & Heap            & Discrete        & Disc           & Rect      & Continuous     & Total           \\ \hline
        25,16h & $59 (7e^{-20})$ & (no data)       & $59 (7e^{-20})$ & (no data)      & (no data) & (no data)      & $59 (7e^{-20})$ \\ 
        26,09h & (no data)       & $72 (1e^{-12})$ & $72 (1e^{-12})$ & $53 (2e^{-2})$ & (no data) & $53 (2e^{-2})$ & $68 (2e^{-13})$ \\ \hline
  Total        & $59 (7e^{-20})$ & $72 (1e^{-12})$ & $62 (2e^{-30})$ & $53 (2e^{-2})$ & (no data) & $53 (2e^{-2})$ & $62 (3e^{-31})$ \\
\end{tabular}
\caption{Finer analysis of the first subject's performance on selecting the maximal value out of four, separated by display mode and combined.
  Note how the average accuracy of random position selection in this case is 25\%, so an accuracy between 53\% and 72\% is a reasonable measure, as well as the probability  to achieve such accuracy or above when choosing one of the value between 4 options uniformly at random}
\label{tab:finerAnalysisTableMaxOutOfFour}
\end{table}

\section{Conclusion}
\label{sec:conclusion}

\begin{HEADPARAGRAPH}
We conclude with a summary of what the project achieved to the date (Section~\ref{sec:achievements}), a discussion of the potential issues with the results presented (Section~\ref{sec:discussion}) and some perspective for future research (Section~\ref{sec:futureWork}).
\end{HEADPARAGRAPH}

\subsection{Achievements}
\label{sec:achievements}

Whereas \citet{2008-AC-TheDiscriminationOfDiscreteAndContinuousAmountsInAfricanGreyParrots-AlAinGiretGrandKreutzerBovet}'s protocol requested the subject to choose between two pieces of cardboard holding distinct amount of food, for discrete and continuous types of food material; we proposed a protocol which requests the subject to choose the largest among a set of values (of parameterized size) on a visual display, using discrete and continuous representations of values, by touching a touchscreen on the representation of the largest value.
By developing a simple but extensively parameterized web application requesting the user to select the largest among two to four values chosen at random, using discrete and continuous representations of values and providing visual and audio feedback about the correctness of the answer, we achieved a solution with various advantages, which we tentatively list as follows.

\subsubsection{Better guarantees against subjects reading potential cues from the experimenter}
\label{maskedTest}
In the context of the measurement of the discrimination abilities between discrete and continuous quantities of subjects, we designed a variant of \citet{2008-AC-TheDiscriminationOfDiscreteAndContinuousAmountsInAfricanGreyParrots-AlAinGiretGrandKreutzerBovet}'s experimental protocol which presents better guarantees against subjects reading potential cues from the experimenter.
Whereas their protocol is performed in presence of a human experimenter who know the complete set-up of the experiment, in our variant the experimenter can ignore the options offered to the subjects and receive audio feedback to indicate whether to reward or not the subject\begin{LONG} (see Section~\ref{maskedExperimentalProtocols} for the definition of a masked experimental set-up)\end{LONG}.

\subsubsection{Generalization of results to  Monk Parakeets}
\label{sec:generalizingResults}

Using such protocol, we replicated and generalized the results obtained by \citet{2008-AC-TheDiscriminationOfDiscreteAndContinuousAmountsInAfricanGreyParrots-AlAinGiretGrandKreutzerBovet} on the discrimination abilities of three African Grey (\emph{Psittacus erithacus}) parrots to that of of two Monk Parakeets (\emph{Myiopsitta Monachus}) parrots.
Concerning the ability to discriminate the largest between 2 values chosen randomly in a domain of 5 distinct values, in discrete or continuous quantities, the two Monk Parakeets parrots performed as well as the three African Grey parrots from \citet{2008-AC-TheDiscriminationOfDiscreteAndContinuousAmountsInAfricanGreyParrots-AlAinGiretGrandKreutzerBovet}'s study, with global accuracies of $82\%$ for the first subject and $74\%$ for the second one (see Section~\ref{sec:maxOutOfTwo} for the detailed results).
Similarly to the results described by \citet{2008-AC-TheDiscriminationOfDiscreteAndContinuousAmountsInAfricanGreyParrots-AlAinGiretGrandKreutzerBovet}, we found a strong correlation between the ratio between the smallest and largest values and the accuracy of the subject: close values are harder to discriminate than others.

\subsubsection{Increased agency of the subject}
\label{sec:increasedAgency}

A subject's sense of \emph{agency}, defined as the faculty for the subject to take decisions and to act upon them, was proven to be an important factor in the well-being of captive nonhuman animals~\cite{2017-TowardsAnAnimalCenteredEthicsForACI,2012-ZB-TechnologyAtTheZooTheInfluenceOfATouchscreenComputerOnOrangutansAndZooVisitors-PerdueClayGaalemaMapleStoinski,1994-RST-ZooAndAnimalWelfare-Kohn}.
In addition to features from the experimental protocol aiming to promote the subject's sense of agency, the web application itself provides various means for the subject to exert its agency, from the ability to choose the mode of display of the values to the ability to interrupt the game at any time and to choose a different mode of display. 

\begin{LONG}
\subsubsection{Extension to tuples}
\label{sec:tuples}

Taking advantage of the extensive parametrization of the web application, we slightly extended the settings of \citet{2008-AC-TheDiscriminationOfDiscreteAndContinuousAmountsInAfricanGreyParrots-AlAinGiretGrandKreutzerBovet}'s study from pairs to tuples: whereas their protocol requested the subject to choose between only two quantities, we were able to study the discrimination abilities not only between pairs of values, but also between triple and quadruple sets of values, showing a reduction of accuracy when the size of such set increased.
\end{LONG}

\begin{LONG}
\subsubsection{Diversifying Discrete and Continuous Representations}
\label{sec:diversifyingRepresentations}

Furthermore, we refined the analysis by diversifying the types of discrete and continuous representations of values (Section~\ref{sec:maxOutOfTwo}), again with the subjects showing an accuracy similar to that of the study of \citet{2008-AC-TheDiscriminationOfDiscreteAndContinuousAmountsInAfricanGreyParrots-AlAinGiretGrandKreutzerBovet}.
\end{LONG}

\subsubsection{Increased Number of Experiments}
\label{sec:increaseNumberOfExperiments}

The web application used in the experiments is similar to other digital life enrichment applications made available to nonhuman animals by their guardians.  Similarly to the behavior described by Washburn~\cite{2015-ABC-TheFourCsOfPsychologicalWellbeingLessonsFrom3DecadesOfComputerBasedEnvironmentalEnrichment-Washburn,1990-BRMIC-TheNASALRCComputerizedTestSystem-RichardsonWashburnHopkinsSavageRumbaughRumbaugh} of apes presented with digital life enrichment applications serving as cognition tests, the subjects often chose to play such web application over other toys made available to them, and often asked to continue playing after the end of a game. This allowed for multiple repetitions of the experiments, and to gather a large amount of data points without incommoding the subjects: the two subjects of this study voluntarily answered a total 2429 tests in one week (see Table \ref{tab:numberOfDataPoints} for a summary), without any observable negative consequences during nor after the end of the testing phase.

\begin{LONG}
\subsubsection{True Randomness}
\label{sec:trueRandomness}

The web application generates the instances presented to the subjects uniformly at random, whereas the high organisational cost of \citet{2008-AC-TheDiscriminationOfDiscreteAndContinuousAmountsInAfricanGreyParrots-AlAinGiretGrandKreutzerBovet}'s protocol limited it to testing the exhaustive enumeration of pairs between values from a specific domain, in a random order.
The later could yield some issues if the domain is sufficiently small that a subject could deduce the answer to some questions by an elimination process, based on previous answers.
As \citet{2008-AC-TheDiscriminationOfDiscreteAndContinuousAmountsInAfricanGreyParrots-AlAinGiretGrandKreutzerBovet} considered a domain of values of size $5$, the amount of distinct unordered pairs is $\frac{5\times4}{2}=10$, a list which subjects with working memory abilities similar to humans might be able to manage.
Beyond the fact that the web application allows the use of a domain of size up to $10$ (which brings the amount of distinct unordered pairs to $\frac{10\times9}{2}=45$), and of sets of values of size larger than two, the fact that the sets of values presented to the subject are generated at random completely suppresses the possibility of a subject to deduce the answer to some questions by an elimination process, based on previous answers.
\end{LONG}

\subsubsection{Automatic generation of the experimental logs}
\label{sec:automaticLog}

The web application automatically generates locally a log of the subject's interactions with it.  This greatly reduces the generation cost of such log, reduces the probability of errors in it, and increases the amount of information captured by it, such as the exact time of each answer, allowing for instance the computation of the amount of time taken to answer each question or studies between the time and/or whether of the day and performance (albeit we did not take advantage of such information in the present study).

\begin{LONG}
\subsubsection{Reduction of Experimental Cost}
\label{sec:costReduction}

As the web application can be run on a simple pocket device, this reduces the cost of running such experiments to the extreme that it can be run on the experimenter's shoulder while the device is hold by hand (at the cost of some accuracy in the results of such experiment).  Such lowered cost might prove to be key in the design of citizen science projects extending this work.
\end{LONG}

\subsection{Discussion}
\label{sec:discussion}

\begin{HEADPARAGRAPH}
Our digital adaptation of \citet{2008-AC-TheDiscriminationOfDiscreteAndContinuousAmountsInAfricanGreyParrots-AlAinGiretGrandKreutzerBovet}'s experimental protocol present some other key difference, which might the result of our study relatively difficult to compare to that of \citet{2008-AC-TheDiscriminationOfDiscreteAndContinuousAmountsInAfricanGreyParrots-AlAinGiretGrandKreutzerBovet}. We attempt to list such difference as follows:
\end{HEADPARAGRAPH}

\subsubsection{Non proportional rewards and reward withdrawal}
\label{sec:nonProportionalRewards}

The protocol defined by \citet{2008-AC-TheDiscriminationOfDiscreteAndContinuousAmountsInAfricanGreyParrots-AlAinGiretGrandKreutzerBovet} instructs to reward the subject with the content of the container they chose: the importance of the reward is proportional to the value being selected. The protocol we defined instructs to reward the subject with a single type of reward each time it does select the maximal value of the set, and to withdraw such reward when the subject fails to do so. Such a difference might alter the experiment in at least two distinct ways:
\begin{itemize}
\item The proportionality of rewards could result in a larger incentive to select the maximal value when the difference between the two values is the largest, and a reduced incentive when the difference is small, and \citet{2008-AC-TheDiscriminationOfDiscreteAndContinuousAmountsInAfricanGreyParrots-AlAinGiretGrandKreutzerBovet} indeed noticed a correlation between the gap between the two values and the accuracy of the answer from the subjects of their experiment. The absence of such proportionality in our experiments might have reduced such an incentive, but we observed the same correlation than they did (described in Section~\ref{sec:relat-betw-perf}).
\item The withdrawal of rewards when the subject fails to select the largest value of the set is likely to affect the motivation of the subject to continue to participate in the exercise on the short term, and in the experiment in the long term. To palliate the frustration caused by such withdrawal, extensive care was taken to progressively increase the difficulty of the exercises (first through the size of the domain from which the values were taken, then through the size of the set of values from which to select the maximal one). No frustration was observed, with both subjects often choosing to continue playing at the end of a game.
\end{itemize}

Implementing the proportionality of rewards is not incompatible with the use of a digital application. For instance, it would be relatively easy to extend the web application to vocalize the value selected by the subject, so that the experimenter could reward the subject with the corresponding amount of food. Such an extension was not implemented mostly because it would slow down the experimentation, for relatively meagre benefits.

\subsubsection{Irregular pairs and tuples}
\label{sec:IrregularPairsAndTuples}

The web application generates the sets of value presented to the subject uniformly at random (without repetitions) from the domain of values set in the parameter page. While such a random generation  yields various advantages, it has a major drawback concerning the statistical analysis of the results, as some sets of value might be under-represented. An unbalanced representation of each possible set of values is guaranteed only on average and for a large number of exercises; whereas \citet{2008-AC-TheDiscriminationOfDiscreteAndContinuousAmountsInAfricanGreyParrots-AlAinGiretGrandKreutzerBovet}'s protocol, using a systematic enumeration of the possible sets of values (presented in a random order to the subject), does not yield such issues.
Such issue was deliberately ignored in order to develop a solution able to measure discrimination abilities on values taken from large domains (assuming that some nonhuman species might display abilities superior to that of humans in this regard), and presenting the subject with a systematic enumeration of the possible sets of values is practical only for small domains (e.g. values from 1 to 5), not for large domains.
For a domain of size $5$ (as that of \citet{2008-AC-TheDiscriminationOfDiscreteAndContinuousAmountsInAfricanGreyParrots-AlAinGiretGrandKreutzerBovet}), enough datapoints were generated that no pair was under represented (see Table~\ref{tab:percentOfCorrecResponsesSubject1}).

\subsubsection{Extension to sensory diverse species}
\label{sec:sensorialDiversity}

The colors displayed by digital displays and the sound frequencies played by devices are optimized for the majority of humans. It is not always clear how much and which colours and sound can be seen and heard by individual of each species. The web application presents extensive parameters to vary the colours displayed and the sounds played to the subject.
Even less intuitively, species can differ in their Critical Flicker Fusion Frequency (CFFF)~\cite{2021-Medicina-CriticalFlickerFusionFrequencyANarrativeReview-MankowskaEtAl}, the frequency at which they perceive the world and can react to it (in some species, such frequency even vary depending on the time of the day or of the season~\cite{2014-SAM-small_animal_live_slow_motion_world-Reas,2013-AB-MetabolicRateAndBodySizeAreLinkedWithPerceptionOfTemporalInformation-HealyMcNallyRuxtonCooperJackson}). For instance, dogs have higher CFFF while cats have lower ones, and the CFFF of reptiles vary with the ambient temperature. Such variation might affect not only their ability to comprehend the visual display and sound play from devices, but might also affect how they comprehend some application designs over others.
The web application presents extensive parameters to vary the time between each exercise and which game, so that part of the rhythm of the application can be adjusted by the experimenter to the CFFF of the subject, but more research is required in order to automatically adapt the rhythm of such applications to the CFFF of individuals from a variety of species.

\subsection{Perspective on future work}
\label{sec:futureWork}

Some issues with the results presented in this work are not related to any difference with \citet{2008-AC-TheDiscriminationOfDiscreteAndContinuousAmountsInAfricanGreyParrots-AlAinGiretGrandKreutzerBovet}'s experimental protocol, but rather with limitations of the current one. We list them along with some tentative solutions, to be implemented in the future.

\subsubsection{Random Dice and Heap representations}
\label{sec:staticRepresentations}

The discrete representation modes \texttt{Dice} and \texttt{Heap} associate each value with a fixed representation of a number of points corresponding to the value being represented. This differs from what happens in \citet{2008-AC-TheDiscriminationOfDiscreteAndContinuousAmountsInAfricanGreyParrots-AlAinGiretGrandKreutzerBovet}'s experimental protocol, where the seeds are in no arranged configuration on the cardboard. This might affect the results of the experience in that a subject could learn to select a particular symbol (e.g. the one corresponding to the largest value of the domain) anytime it is present, without any need for any comparison between the presented values.
\begin{MAYB}
Check in the results if value sets including the largest value of the domain have a better accuracy ratio than others: this could be an indication that the subjects learned to select the corresponding symbol anytime it is present, without any need for comparing values.
\end{MAYB}
The development and evaluation of their impact on the discrimination abilities of human and nonhuman subjects will be the topic of a future study, once the corresponding randomized representations have been added to the web application.
\begin{JeremySays}FABIAN: should you want to work on a second article after this one, this topic might be either to study than the campaign mode, and the random display should be quick enough to program...\end{JeremySays}

\subsubsection{Systematic logs}

The easiness with which logs are generated tends to make one forget about it, to the point that the bottleneck could become the transfer of the logs from the device used to perform the experience to a central repository. As one guardian might get more excited to transfer the logs of sessions where the subjects excelled at the activities than that of less positive sessions, this might create a bias toward positive results in their report. While not an issue while implemented by personal with a scientific training, such risk of a bias might become more problematic in the context of a citizen science project~\cite{Website-CitizenScience}.
The development of a website serving as a central repository of experimental data sent by web applications such as the one presented in this work will be the topic of a future study. The roles of such a central ``back-end'' website could include
  the automatizing of the most frequent statistical tests on the data received; 
  a greater ease of separation between the roles of experimenter and researcher, which will be an important step toward a true citizen science generalisation of this project; and 
  the aggregation of sensory and cognitive data from distinct applications, individuals and species.

\subsubsection{Adaptive Difficulty}
\label{sec:complexDifficultySettings}

The great amount of parameters available in the settings page of the web application makes it possible to adapt the difficulty of the activities to the level of abilities of the subject.
Such abilities evolve with time, most often advancing and only rarely receding (such as after a long period without using the web application).
Choosing which values of the parameters is the most adequate to the current level of abilities of the subject requires an extensive understanding of the mechanisms of the application.
An extension of the web application presenting the subject with a sequence of parametrization of increasing difficulty, along with a mechanism raising or lowering the difficulty of the activities presented to the subject would greatly simplify the task of the experimenter, and will be the topic of a future study.

\begin{LONG}
\subsubsection{Cardinal Discrimination}
\label{sec:cardinalDiscrimination}

\citet{2006-CP-OrdinalityAndInferentialAbilitiesOfAGrreyParrot-Pepperberg} recounts how the African Grey parrot (\emph{Psittacus erithacus}) Alex, after being trained to identify Arabic numerals from 1 to 6 (but not to associate Arabic numbers with their relevant physical quantities) was able to label which of two Arabic numerals is the biggest, having inferred the relationship between the Arabic number and the quantity, and having understood the ordinal relationship of his numbers.
%
%
Modifying the web application\begin{NONANONYMOUS}\texttt{InCA-WhatIsMore}\end{NONANONYMOUS} so that to replace the graphical representations of values by ordinal numbers would/will be easy. Testing ethically the ability or inability of subjects to replicate \citet{2006-CP-OrdinalityAndInferentialAbilitiesOfAGrreyParrot-Pepperberg}'s results without frustrating those subjects might require more sophistication in the design of the experimental protocol.
Such protocols concerning the measurement of skills that subject might lack is the topic of the next section.
\end{LONG}

\begin{LONG}
\subsubsection{Ethical Measurement of Inabilities}
\label{sec:EthicalMeasurementOfUNabilities}

The frustration potentially caused by the  withdrawal of rewards (described in Section~\ref{sec:nonProportionalRewards}) when measuring skills that a subject might lack (an example of which was given in Section~\ref{sec:nonProportionalRewards}) points out to another issue, of ethical dimensions: how can one ethically demonstrate the inability of subjects to perform a given action through experimentation, without hurting the well-being of the subject by exposing them to the frustration of failing to performed the requested action?
Note that such issue is not specific to the action of withdrawing rewards when a subject fails: as proportional rewards can also generate frustration. 
One solution could be to mix potentially ``difficult'' requests with other, similar but known to be ``easy'', requests, in such a way that the proportion and frequency of the former to be a fraction of the proportion and frequency of ``easy'' requests that the subject fail (for inattention or other reasons). One can hypothesize that 1) the frustration generated by such questions would be minimum; that 2) a statistical analysis of the correction of the difficult requests would yield useful information about the ability or inability of the subject to answer those; and that 3)  a small proportion of ``difficult'' requests helps to further motivate the subject, making the exercise more of a challenge.
\end{LONG}

\begin{LONG}
\subsubsection{Citizen Science Extensions}
\label{sec:citizenScienceProject}

The term ``\emph{Citizen Science}'' refers to scientific projects conducted, in whole or in part, by amateur (or nonprofessional) scientists~\cite{2013-NATURE-CitizenScienceAmateurExperts-Gura}.
It is sometimes described as ``public participation in scientific research'', with the dual objectives to improve the scientific community's capacity, as well as improving the public's understanding of science and conscience about the research's themes~\cite{Website-CitizenScience}.  Citizen Science has become a means of encouraging curiosity and greater understanding of science whilst providing an unprecedented engagement between professional scientists and the general public.

Such methodology must be used with care, in particular about the validity of volunteer generated data.
Projects using complex research methods or requiring a lot of repetitive work may not be suitable for volunteers, and the lack of proper training in research and monitoring protocols in participants might introduce bias into the data~\cite{2008-PS-CultivatingConnectionIncorporatingMeaningfulCitizenScienceIntoCapeCodNationalSeashoreEstuarineResearchAndMonitoringPrograms-ThelenThiet}.
Nevertheless, in many cases the low cost per observation can compensate for the lack of accuracy of the resulting data~\cite{2012-FEE-LessonsFromLadyBeetles-GardinerEtAl}, especially if using proper data processing methods~\cite{2020-Patterns-ArtificialIntelligenceMeetsCitizenScienceToSuperchargeEcologicalMonitoring-McClureEtAl}.

Scientific researchers in comparative psychology could definitely benefit from some help, with many cognitive aspects to explores for so many species.
In the process of defining the \emph{anecdotal method} of investigation for creative and cognitive processes, \citet{2007-Methods-CreativeOrCreatedUsingAnecdotesToInvestigateAnimalCognition-BatesByrne} mentioned that ``\emph{collation of records of rare events into data-sets can illustrate much about animal behaviour and cognition}''. Now that the technology is ready to analyze extremely large data-sets, what is lacking in comparative psychology are the means to gather such large data-sets.

Delegating part of the experimental process to citizens without proper scientific training is not without risk.
Given the conflicted history of Comparative Psychology~\cite{2020-FiP-TheComparativePsychologyOfIntelligenceSomeThirtyYearsLater-Pepperberg} in general and Animal Language Studies~\cite{2017-Psychonomic-AnimalLanguageStudiesWhatHappened-Pepperberg} in particular, the challenge of avoiding ``Clever Hans'' biases and related ones will be of tremendous importance.
%
Could applications and experimental protocols such as the one described in this work help to design citizen science projects for the study of sensory and cognitive abilities in nonhumans species living in close contact with humans?
\end{LONG}


\begin{contrib}
Jérémy Barbay programmed the first versions of the software, managed the interactions with the subjects during the development, training and testing phases, obtained the approval of the \emph{Institutional Animal Care and Use Committee}, structured the article and supervised the work of Fabi\'an Jaña Ubal and Cristóbal Sepulveda Álvarez.
Fabi\'an Jaña Ubal improved and maintained the software, and described it (Sections~\ref{sec:applicationStructure} and~\ref{sec:maskedExperimentalSetup}).
Cristóbal Sepulveda Álvarez reviewed the experimental results, performed their statistical analysis, and described the log structure (Section~\ref{sec:loggingStructure}), the statistical analysis process (Section~\ref{sec:statisticalAnalysis}) and the results (Section~\ref{sec:results}).
All authors are aware of the submission of this work and of its content.
\end{contrib}

\begin{acks}
We wish to thank 
Joachim Barbay for his suggestion of using \texttt{Svelte} and his mentoring in the development of the various versions of the application;
Jennifer Cunha from \textsc{Parrot Kindergarten} for sharing images and video of parrots using touchscreens, suggesting the \cite{2008-AC-TheDiscriminationOfDiscreteAndContinuousAmountsInAfricanGreyParrots-AlAinGiretGrandKreutzerBovet}'s study and for helping during the design and test of the preliminary versions of the software \texttt{InCA-WhatIsMore} (as well as other software projects);
    Corinne Renguette for her help concerning the bibliography and the ethical aspects of the experiments and of its description; 
    Cristina Doelling for pointing out some of the existing literature about the use of touchscreens by apes in zoo; and 
    Francisco Gutierrez and Jenny Stamm for suggesting  alternative names to the problematic term ``blind'' in expressions such as ``blind setup'', and for pointing out some bibliography supporting such replacement.
\end{acks}

\bibliographystyle{ACM-Reference-Format}
\interlinepenalty=10000
\bibliography{aci}



\end{document}